\documentclass[pra,aps,tightenlines,twocolumn,showpacs,amssymb,amsfonts,superscriptaddress,eqsecnum]{revtex4}

\newcommand{\ket}{\rangle}
\newcommand{\bra}{\langle}

\usepackage{epsfig}

\begin{document}
\title{Adaptive phase estimation is more accurate than non-adaptive phase estimation for continuous
beams of light}
\author{D. T. Pope}
\email{d.pope@griffith.edu.au} \affiliation{Centre for Quantum
Dynamics, School of Science, Griffith University, Brisbane 4111,
Queensland, Australia}
\author{H. M. Wiseman}
\email{h.wiseman@griffith.edu.au} \affiliation{Centre for Quantum
Dynamics, School of Science, Griffith University, Brisbane 4111,
Queensland, Australia} \affiliation{Centre for Quantum Computer
Technology, Griffith University, Brisbane 4111, Australia}
\author{N. K. Langford}
\email{langford@physics.uq.edu.au} \affiliation{Centre for Quantum
Computer Technology, \\ School of Physical Sciences, University of
Queensland, Brisbane 4072, Queensland, Australia}

\date{\today}

\begin{abstract}
We consider the task of estimating the randomly fluctuating phase
of a continuous-wave beam of light. Using the theory of quantum
parameter estimation, we show that this can be done more
accurately when feedback is used (adaptive phase estimation) than
by {\em any} scheme not involving feedback (non-adaptive phase
estimation) in which the beam is measured as it arrives at the
detector. Such schemes not involving feedback include all those
based on heterodyne detection or instantaneous canonical phase
measurements.
%for which the field is measured via heterodyne
%detection. Furthermore, we show that in this scenario heterodyne
%detection is the best non-adaptive detection method in which the
%field is measured in real time.
%It follows
%that adaptive phase estimation is more accurate than non-adaptive
%phase estimation
We also demonstrate that the superior accuracy adaptive phase
estimation is present in a regime conducive to observing it
experimentally.
%whilst adaptive is more
%accurate than real-time non-adaptive phase estimation for {\em
%all} photon fluxes it is more accurate than it for small to
%moderate photon fluxes, the regime that is most experimentally
%conducive to seeing a difference between the two schemes.
%{\em (Possible alternate
%last sentence: We also show that, most probably, it is better than
%any non-adaptive estimation scheme at all.)}
\end{abstract}


\maketitle

\section{Introduction} \label{intro}

%
%gently introduce the topic
%

Phase is a physical property found in both classical and quantum
electromagnetic (EM) fields. For classical EM fields comprising a
single mode, it can be determined exactly via measuring two
orthogonal quadratures or components of such fields. This,
however, is not the case for single-mode EM fields in quantum
mechanics.
%quantum-mechanical single-mode EM field
%For instance, the phase of the coherent
%state $||\alpha| e^{i \phi}\ket$ is $\phi$ which is a state
%parameter. Importantly, $\phi$ is {\em not} an observable. For
%instance, it is not the observable associated with the
%Pegg-Barnett phase operator \cite{pegg88} which is also called
%phase but which is something different and does not have a
%well-defined value for $||\alpha| e^{i \phi}\ket$.
%Though we can exactly determine phase classically, this is often
%not the case in quantum mechanics.
Estimates of the phases of such fields are necessarily imperfect
due to intrinsic quantum noise in measurements of non-commuting
observables such as quadratures. Given this limitation, quantum
{\em phase estimation}, the process of estimating the phase of a
quantum-mechanical EM field as accurately as possible, is
non-trivial.

%
%phase estimaton is interesting
%

In addition to being non-trivial, phase estimation in quantum
mechanics is interesting for a number of reasons. First, at some
time in the future it may be practical to encode and send
information in the phase of a single electromagnetic field mode at
or near the ultimate quantum limit --- the upper limit permitted
by quantum mechanics \cite{hall91,caves94,mabuchi04}. In such a
scenario, the more accurately a receiver could estimate phase the
more information could be sent. Second, it also may be useful in
interferometric gravity-wave detection.
%Third, phase estimation is
%interesting for fundamental reasons. Its accuracy is necessarily
%limited by intrinsic quantum noise and so it is informative to
%consider how well can we estimate phase in the face of this
%unavoidable noise.
Third, phase estimation is interesting as it is an instance of
quantum parameter estimation \cite{verstraete01,wiseman95}, an
increasingly experimentally accessible field concerned with
estimating parameters of quantum states as well as possible in the
face of unavoidable quantum noise.
%Indeed, a recent experiment
%\cite{armen02} estimated the phase of a single pulse of light
%using a highly accurate estimation technique \cite{wiseman95}.

%
%note for us phase is a parameter, not an observable
%

%It is important to note that throughout this paper by the term
%``phase" we mean a particular state parameter and not an
%observable. For instance, consider the coherent state $||\alpha|
%e^{i \phi}\ket$. For us, its phase is the parameter $\phi$, which
%has a well-defined value. It is not, for example, the observable
%associated with the Pegg-Barnett phase operator \cite{pegg88}
%which is also called phase but which is something different and
%does not have a well-defined value for $||\alpha| e^{i \phi}\ket$.
%As a result, phase estimation is an example of quantum parameter
%estimation, albeit sometimes with a time-varying parameter.

%
%two types of phase estimation: adaptive versus non-adaptive
%

Phase can be estimated via two broad approaches, {\em
non-adaptive} phase estimation and {\em adaptive} phase estimation
\cite{wiseman95}. In non-adaptive phase estimation, which is the
conventional approach, we measure an EM field via a single fixed
measurement that remains constant over time. In adaptive phase
estimation, however, the measurement is continually adjusted in an
attempt to maximise its accuracy at each moment in time. This is
done by changing or {\em adapting} it based on earlier measurement
results. For both EM-field pulses and also continuous EM beams, it
has been shown that adaptive phase estimation is more accurate
than (at least) many instances of the conventional non-adaptive
approach \cite{wiseman95,wiseman97,wiseman98,berry02}.

%
%outline of the paper
%

In this paper we consider the problem of estimating the randomly
fluctuating phase of a continuous-wave (cw) EM field (EM beam) as
introduced in Ref.~\cite{berry02}. We show that this can be done
more accurately using adaptive phase estimation than via {\em any}
non-adaptive phase estimation scheme in which the field is
measured in real time (that is, as it arrives at the detector). We
also show that this improved accuracy exists for fields with small
to moderate photon fluxes. These are our two main results. The
latter is significant, first, as a {\em theoretical} difference
between the accuracies of adaptive and non-adaptive phase
estimation is most readily seen {\em experimentally} in fields
with small to moderate photon fluxes. Second, in a communication
scenario in which a receiver is trying extract information encoded
in the phase of an EM field by a distant sender, it is likely that
the receiver will be making measurements on fields with small to
moderate photon fluxes due to attentuation \cite{armen02}. In the
course of arriving at the two results, we present a theoretical
technique for estimating phase that may be applicable to a range
of problems. Our results build upon earlier work
\cite{wiseman95,wiseman97,wiseman98,berry02} and further
demonstrate the superiority of adaptive scheme over conventional
non-adaptive ones for the important task of phase estimation.
%They
%also suggest that for cw EM fields we may be able to see this
%superiority experimentally.

%
%structure of the paper
%

This paper is structured as follows: In Section~\ref{theory_sec},
we review the mathematical tools used throughout. They are Bayes'
rule, the Kushner-Stratonovitch equation and the Zakai equation.
Next, Section~\ref{new_theory_sec} presents the phase estimation
schemes considered, some of which are adaptive and some of which
are non-adaptive. In Section~\ref{results_sec}, we compare the
accuracies of the schemes in the steady-state regime, showing that
each of the adaptive schemes is more accurate than all of the
non-adaptive schemes.
%In particular, we show that the
%adaptive schemes are more accurate than the non-adaptives for
%small to moderate photon fluxes, the regime in which a difference
%is most likely to been experimentally observed.
Finally, in
Section~\ref{discussion} we discuss our results.

%
%work already done
%

Before proceeding further, we first review existing work on
adaptive phase estimation. As previously stated, the conventional
method for estimated the phase of an EM field is via non-adaptive
phase estimation. For a single-mode EM-field {\em pulse} in the
coherent state $|\beta  \ket$, where $\beta \in {\mathbb C}$, the
most widely known method \cite{caves94,armen02} of estimating the
phase $\phi \left( ={\rm arg}(\beta)\right)$ uses a non-adaptive
detection technique called heterodyne detection
\cite{yuen78,yuen80,shapiro79,shapiro84,shapiro85,caves94}. This
involves mixing the pulse, which we call the {\em signal pulse},
with an intense local oscillator of phase $\Phi=\Phi_{0} +\Delta
t$ at a 50:50 beamsplitter. Here $\Delta$ is a detuning, $t$
denotes time and $\Phi_{0}$ is the phase at $t=0$. The difference
between the photocurrents in the beamsplitter's two output ports
is proportional to the quadrature phase amplitude $X_{\Phi}=a
e^{-i\Phi}+ a^{\dag} e ^{i \Phi}$, where $a$ and $a^{\dag}$ are
creation and annihilation operators for the signal pulse. Assuming
that $\Delta \gg \Gamma$, where $\Gamma$ is the signal pulse's
spectral width, all quadratures are rapidly measured and thus, for
all practical purposes, heterodyne detection instantaneously
measures the complex photocurrent $I_{\rm c}$ containing equal
information about the observables $X_{\Phi=0}$ and
$X_{\Phi=\pi/2}$. Once the signal pulse has been measured, $\phi$
can then estimated from an appropriate functional of all the
recorded currents. For large values of $|\beta|$ this approach
leads to an estimate with a variance of $1/\left( 2 |\beta|^{2}
\right)$ \cite{wiseman97}. Half of this is non-fundamental and
results from excess noise introduced by heterodyne detection due
to the fact that it measures two noncommuting quadratures. This
excess contribution to the variance can also be thought of as
arising from the fact that heterodyne detection measures {\em all}
quadratures equally. Because of this, it sometimes measures some
the so-called amplitude quadrature ($X_{\Phi=\phi}$) which
contains {\em no} information about $\phi$.

%
%the cw scenario
%

A second type of EM field for which phase estimation has been
considered is a {\em continuous} EM beam. In particular, Ref.
\cite{berry02} considered such estimation for a continuous beam in
a coherent state with phase $\phi$ that
%$\alpha(t)(=|\alpha(t)|e^{i\phi(t)})$ such that $|\alpha(t)|^{2}$
%is constant and equal to the mean number of photons per unit time
%in the beam
randomly fluctuated in time as a Wiener process \cite{gardiner83}.
This paper found that one particular {\em non-adaptive} phase
estimation scheme estimated $\phi$ with a variance of
$1/\sqrt{2N}$ in the steady-state regime for $N \gg 1$. Here, $N$
is the beam's photon flux in an amount of time equal to its
coherence time (which is set by the timescale of the fluctuations
in $\phi$).

%
%adaptive phase est. in the single-shot scenario
%

\begin{figure}
\center{\epsfig{figure=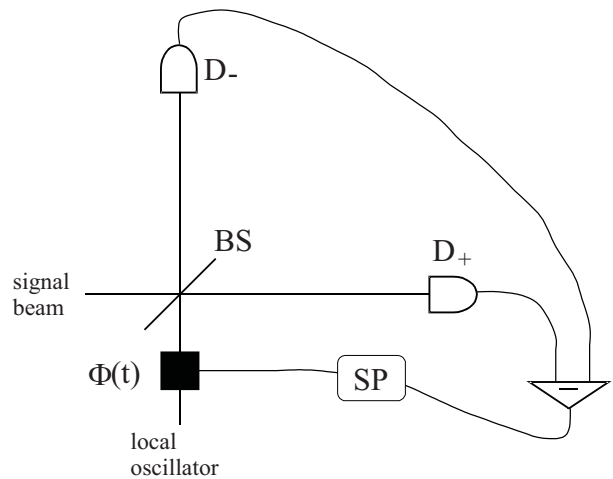,width=80mm}}
\caption{Schematic diagram of the measurement set-up for adaptive
homodyne-based phase-estimation schemes. The symbol BS denotes the
50:50 beamsplitter, $D_{-}$ and $D_{+}$ are photon counters for
which the difference in the number of photons they detect is found
and then fed back to the local oscillator's phase. A signal
processor is denoted by $SP$.} \label{figure_adaptive_diag}
\end{figure}

Though non-adaptive phase estimation using heterodyne detection
yields a reasonable estimate of $\phi$ for both a single EM-field
pulse and a continuous EM beam, this quantity can be more
accurately estimated via adaptive techiques \cite{wiseman95,
wiseman97, wiseman98, berry02}. For a single pulse of light, again
in the coherent state $| \beta \ket$, this can be done by
measuring the field using adaptive homodyne detection.
Non-adaptive homodyne detection is identical to heterodyne
detection except that the local oscillator has the same frequency
as the pulse's mean frequency so that $\Phi$ is constant
\cite{mandel}. It is made adaptive by varying $\Phi$ so as to try
to measure the so-called phase quadrature. This is the quadrature
for which $\Phi = \phi + \pi/2$, and, moreover, the one that
minimises the measurement's excess uncertainty, below that of
heterodyne detection. To try to measure the phase quadrature we
use the results of previous measurements to obtain ${\hat
\phi}_{\rm fb}(t)$, an estimate for $\phi(t)$. This is then fed
back to the local oscillator and $\Phi$ is set to $\Phi(t)={\hat
\phi}_{\rm fb}(t)+ \pi/2$ in an attempt to `home in' on the phase
quadrature. Fig.~\ref{figure_adaptive_diag} shows a schematic
diagram of the apparatus implementing this scheme.
When $|\beta|$ is large, it leads to a variance in our estimate of
$1/\left(4 |\beta|^{2} \right)$ \cite{wiseman97}, which is only
half as large as that of the non-adaptive scheme discussed above.
Furthermore, this improved accuracy has been seen experimentally
\cite{armen02}.

For the continuous EM beam with a randomly fluctuating phase
considered earlier, it is known that a particular adaptive scheme
is more accurate than one particular non-adaptive one
\cite{berry02}. But is it also more accurate than the best
possible non-adaptive scheme? One of main results of this paper is
to show, in Section~\ref{results_sec}, that in the steady-state
regime adaptive phase estimation is more accurate than {\em any}
non-adaptive estimation scheme in which the EM field is measured
in real time, even one involving a canonical phase measurement
\cite{leonhardt95}. In addition, we show that the improved
accuracy of adaptive phase estimation persists for $N \ll 1$.

\section{Background Theory} \label{theory_sec}

\subsection{What is phase?}
Within quantum mechanics, the term `phase' has multiple meanings
\cite{barnett93,leonhardt95}. In this paper, however, it refers to
a single concept which we now state. The electric field of a {\em
classical} single-mode EM-field pulse incident on an ideal
photodetector is, in the vicinity of this detector,
\begin{equation} \label{em_field}
\vec{E}(t)=\sqrt{\frac{2\hbar \omega}{\epsilon_{0}{\cal A} c}}
\vec{e} \left( |\alpha| e^{ - i( \omega t - \phi_{\rm cl.})} +
{\rm c.c.} \right).
\end{equation}
Here, $t$ denotes time, $\omega$ is the field's angular frequency,
$\epsilon_{0}$ denotes the permittivity of free space, ${\cal A}$
is the transverse area over which the field is spread, $\vec{e}$
represents a unit vector denoting the field's direction,
$|\alpha|$ is a complex amplitude with dimensions of ${\rm
time}^{-1/2}$, $c$ denotes the speed of light and c.c. represents
a complex conjugate.
%The
%$E$ of the pulse after the detector is zero and the $E(time=T)$ at
%point $x$ is $E_{detector}(t+T)$.
Given this, we define $\phi_{\rm cl.}$ to be this field's phase.
Similarly, the phase of a quantum-mechanical single-mode EM-field
pulse is defined to be the quantum-mechanical analogue of
$\phi_{\rm cl.}$, which we denote by $\phi$. For instance, the
phase of the coherent state $||\beta| e^{i \phi}\ket$ is defined
to be $\phi$ which is a parameter and {\em not} an observable. In
particular, it is not the observable associated with the
Pegg-Barnett phase operator \cite{pegg88} which is also called
phase but which does not have a well-defined value for the state
$||\beta| e^{i \phi}\ket$.

\subsection{Continuous EM beam}
The scenario which we consider throughout this paper
%We now describe in detail the scenario within which we
%measure the phase of a continuous EM beam \cite{berry02}. It
centres around a continuous EM beam \cite{berry02} known as the
{\em signal beam}. This beam is the output of an idealised laser,
and so can be described by a coherent state with complex amplitude
$\alpha$. The mean photon flux $|\alpha|^{2}$ is constant.
However, the beam's phase $\phi(t)$ fluctuates randomly such that,
again in the vicinity of the detector,
\begin{equation} \label{true_phase}
\frac{d \phi}{dt} = \sqrt{\kappa} \xi(t).
\end{equation}
Here $\kappa$ is a noise strength and $\xi$ is real Gaussian white
noise defined by
\begin{equation}
\bra \xi(t) \xi(t') \ket = \delta( t- t').
\end{equation}
In practice, this fluctuation could be achieved via an
electro-optical modulator (EOM) \cite{yariv89} that `imprints' a
fluctuating phase on each segment of the beam. These phase
fluctuations give the beam a linewidth of $\kappa$, so that
$N=|\alpha|^{2}/\kappa$ is the number of photons in the coherence
time.

\begin{figure}
\center{\epsfig{figure=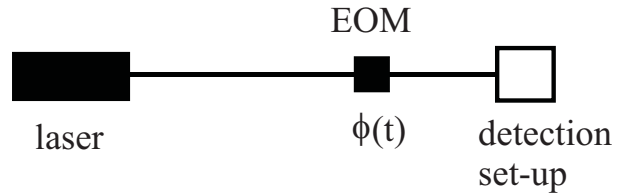,width=80mm}}
\caption{Schematic diagram of the physical scenario considered.
First, an idealised cw laser outputs a continuous beam of light
which is then incident on an electro-optical modulator (EOM). The
EOM imprints phases on segments of the beam which are then
incident on the detection set-up on the right.} \label{fig_beam}
\end{figure}

%
%the measurement model for the cw scenario
%

%We now describe the nature of photodetection in the cw scenario.
In the continuous EM beam scenario, we measure the signal beam via
either homodyne or heterodyne detection. For homodyne detection,
the photocurrent $I_{\rm r}$ measured in the interval $dt$ is
given by
\begin{equation} \label{homo_measurement}
I_{\rm r} dt = 2 \eta {\rm Re} \left( |\alpha| e^{i(\phi-\Phi)}
\right)dt + \sqrt{\eta} dW.
\end{equation}
Here $dW$ is a real Wiener increment, $\eta$ is the detector's
efficiency (which is its probability of detecting an incident
photon) and $\Phi$ is the local oscillator phase. In contrast,
heterodyne detection simultaneously measures the quadratures
$X_{\Phi=0}$ and $X_{\Phi=\pi/2}$. An alternate way of doing this
is to first split the signal beam at a 50:50 beamsplitter and then
to measure $X_{\Phi=0}$ at one output and $X_{\Phi=\pi/2}$ at the
other. Assuming perfect detectors, each photodetector measures, on
average, half of the beam's photons and thus the quantum
efficiency of each measurement is $\eta=1/2$. Representing both
outcomes in terms of a single complex quantity, we obtain
\begin{equation} \label{hetero_measurement}
I_{\rm c} dt =|\alpha| e^{i \phi} dt + dW_{\rm c},
\end{equation}
where $dW{\rm c}$ is a complex Wiener increment defined by the
correlations $\bra dW_{\rm c}  dW^{*}_{\rm c}\ket=dt$ and $\bra
dW_{\rm c} dW_{c} \ket = 0$.

\subsection{Non-adaptive and adaptive phase estimation}
\label{parameter_estimation_theory} In a number of the
phase-estimation schemes we consider, $\phi$ is estimated using
the theory of quantum parameter estimation
\cite{verstraete01,wiseman95}. This process involves two steps.
First, Bayes' rule is used to obtain a differential equation with
respect to time for $P(\phi)$, the probability distribution
encoding our knowledge of $\phi$, which we then solve. Bayes' rule
updates our knowledge of some unknown parameter given the
measurement result $M$. For the situations we consider, it is
\begin{equation} \label{bayes_rule}
P(\phi|M) = \frac{P(\phi) P(M|\phi)}{P(M)},
\end{equation}
where $P(x|y)$ denotes the probability of $x$ given $y$. The
second step in the process of estimating $\phi$ via quantum
parameter estimation is to use $P(\phi)$ to calculate our estimate
of $\phi$, which we denote by ${\hat \phi}(t)$.

%
%how we use Bayes rule in general
%

To explain in more detail the first step of generating and solving
a differential equation for $P(\phi)$, we begin by observing that
in Eq.~(\ref{bayes_rule}) the term $P(M)$ is a normalisation
factor that ensures the normalisation of $P(\phi|M)$. This can be
seen by realising that we can write $P(M)$ as
\begin{equation}
P(M) = \int_{\phi=\phi_{0}}^{\phi_{0}+ 2 \pi} P(\phi) P(M|\phi)
d\phi,
\end{equation}
where $\phi_{0}$ is an arbitrary lower limit. It follows from this
that upon replacing $P(M)$ in Eq.~(\ref{bayes_rule}) by another
function of $M$ that is independent of $\phi$ we obtain a
quasi-Bayes' rule that updates an {\em unnormalised} `probability'
distribution for $\phi$ that we label $\tilde{P}(\phi)$
\cite{verstraete01}. We choose to replace $P(M)$ by
$P(M)_{|\alpha|=0}$, where $P(M)_{|\alpha|=0}$ is the probability
of measuring the result $M$ given $|\alpha|=0$, and so
Eq.~(\ref{bayes_rule}) becomes
\begin{equation} \label{quasi_bayes_rule}
\tilde{P}(\phi|M) = \frac{\tilde{P}(\phi)
P(M|\phi)}{P(M)_{|\alpha|=0}}.
\end{equation}
The function $P(M)_{|\alpha|=0}$ was chosen as it corresponds to
considering the measurement result $M$ in the denominator to be
Gaussian white noise which, in turn, simplifies
Eq.~(\ref{bayes_rule}). Furthermore, it yields a liner evolution
equation for $\tilde{P}(\phi)$. This is in contrast to the
non-liner one for $P(\phi)$ that would have been obtained had
$P(M)$ not been replaced.

%
%getting from the quasi-Bayes rule to the Zakai eqn and the KSE eqn
%

The next step in obtaining and solving a differential equation for
$P(\phi)$ is to transform Eq.~(\ref{quasi_bayes_rule}) into the
form
\begin{equation} \label{zakai_eq}
d\tilde{P}(\phi) =  \left( f(\phi) g(M) + {\rm c.c.} \right)
\tilde{P}(\phi) dt,
\end{equation}
where $f(\phi)$ and $g(M)$ are functions whose nature depends upon
$P(M|\phi)$ and $P(M)_{|\alpha|=0}$, by neglecting terms of order
$dt^{2}$ or higher. This equation is known as a {\em Zakai
equation} \cite{benssousan92}. To obtain the desired differential
equation for $P(\phi)$ with respect to time from
Eq.~(\ref{zakai_eq}) we normalise $\tilde{P}(\phi)$ using a known
procedure detailed in Appendix A. This leads to the following
differential equation for $P(\phi)$:
\begin{eqnarray} \label{kse} \nonumber
d P (\phi) & = & |\alpha | \left[(e^{i\phi} -\bra e^{i\phi}
\ket_{P(\phi)}) P(\phi) \zeta(t)  + {\rm c.c.} \right]dt, \\
\end{eqnarray}
where $\zeta$ is either real or complex Gaussian white noise
depending on the nature of $M$. This is known as a {\em
Kushner-Stratonovitch (KS) equation} \cite{mcgarty74}.

%
%adding the phase-diffusion term
%

Thus far, we have only considered the evolution of $P(\phi)$ due
to our measurement of the signal beam. However, there is also its
evolution resulting from the diffusion described by
Eq.~(\ref{true_phase}). In the absence of measurement, this
diffusive evolution leads to $P(\phi)$ being a Gaussian
distribution centred on $\phi(t=0)$ with variance $\kappa t$. A
straightforward calculation shows that the evolution equation for
this distribution in this case is the Fokker-Planck equation
\begin{equation}
dP(\phi) = \frac{\kappa}{2} \frac{\partial^{2}P(\phi)}{\partial
\phi^{2}}dt.
\end{equation}
Adding the effects of phase diffusion to Eq.~(\ref{het_kse}) leads
to the final KS equation
%\begin{eqnarray} \label{kse_two}
%d P & = & \frac{\kappa}{2} \frac{\partial^{2} P } {d \phi^{2}}dt
%\\ \nonumber & + & |\alpha | \left[(e^{i\phi} -\bra
%e^{i\phi} \ket) P \zeta(t)  + {\rm c.c.} \right]dt.
%\end{eqnarray}
\begin{eqnarray} \label{kse_two} \nonumber
d P & = & \frac{\kappa}{2} \frac{\partial^{2} P } {d \phi^{2}}dt +
|\alpha | \left[(e^{i\phi} -\bra e^{i\phi} \ket_{P(\phi)} P
\zeta(t) +
{\rm c.c.} \right]dt. \\
\end{eqnarray}
Solving this equation we obtain $P(\phi)$.
%Eq.~(\ref{kse_two}) can
%be viewed as a Fokker-Planck equation to which we have added the
%effects of measurement via the second term on the right-hand side.

%
%getting the estimate from the KSE
%

As stated at the start of this subsection, the second step in
estimating $\phi(t)$ via quantum parameter estimation is to
calculate the optimal estimate for $\phi(t)$ from $P(\phi)$. This
is defined to be the estimate with the following two properties:
\begin{enumerate}
\item It has the smallest possible average error as measured by
the Holevo variance \cite{holevo84}. \item It is such that $ \bra
\exp[i(\phi -{\hat \phi})] \ket_{I,\xi} \in {\mathbb R}_{+}$. Here
$\bra \ldots \ket_{I,\xi}$ is an average over $I$ and $\xi$, where
$I$ is either $I_{c}$ or $I_{\rm r}$ depending on the measurement
scheme.
\end{enumerate}
The Holevo variance is a measure of statistical spread suitable
for any cyclical variable $x$ and is given by
\begin{equation} \label{holevo_define}
V^{H}(x) = |\bra e^{ix} \ket|^{-2}-1.
\end{equation}
For such variables, it is superior to the standard variance
$\sigma^{2}$ as the latter can be ill-defined. To illustrate this
problem, observe that $\phi$ has the range
$[\phi_{0},\phi_{0}+2\pi)$, where $\phi_{0}$ is usually chosen to
be either $-\pi$ or $0$. As a result, depending on our choice of
$\phi_{0}$, $\sigma^{2}(\phi)$ can take different values for a
single distribution. The reason for the second property is to rule
out estimates with small Holevo variances but which are
systematically biased and hence do not estimate $\phi$ accurately.
%that an
%estimate may have a small Holevo variance but yet still be a bad
%estimate. This occurs when ${\hat \phi}$ is of the form ${\hat
%\phi} = \phi +\epsilon (\xi, \eta) + c$, where $\epsilon$ is a
%fluctuating offset such that $\epsilon \in {\mathbb R}$ and
%$|\epsilon| \ll 1$, and $c$ is a constant offset. When ${\hat
%\phi}$ can be written in this manner then it has a large error but
%yet only a small fraction of this error fluctuates stochastically.
%as $\xi$ and $\eta$
%fluctuate.
%As the Holevo variance of the error $V^{H}(\phi -{\hat \phi})$
%only depends on the magnitude of $\bra \exp[i(\phi - {\hat
%\phi})]\ket_{\xi,\eta}$ it follows that it is independent of $c$.
%As a result, $c$ can be large but yet $V^{H}(\phi -{\hat \phi})$
%can still be small. To avoid considering the estimates in such
%scenarios as optimal, we demand that the optimal estimate
%satisfies the second property. This property only satisfied in
%situations such as when the argument of the mean of $e^{i{\hat
%\phi}}$ (averaged over $\eta$) is $\phi$ and thus that ${\hat
%\phi}$ is centred on $\phi$. Both properties together are only
%satisfied by estimates whose fluctuations are as small as possible
%{\em and} that are centred on $\phi$.
%Requiring the satisfaction
%of  and thus that ${\hat \phi} not only possesses small
%fluctuations (and thus leads to $V^{H}(\phi -{\hat \phi})$ being
%small) but also

The optimal estimate we wish to calculate is given by
\begin{equation} \label{opt_estimate}
{\hat \phi}(t) = {\rm arg} \left( \bra \exp(i\phi) \ket_{P(\phi)}
\right),
\end{equation}
where $\bra \ldots \ket_{P(\phi)}$ denotes an average over
$P(\phi)$. Whilst the estimate $\bra \phi \ket_{P(\phi)}$ is a
more obvious choice for the optimal estimate of $\phi(t)$, it
sometimes estimates $\phi(t)$ poorly due to the fact that
$\phi(t)$ is a cyclical variable. This occurs, for instance, when
$P(\phi)$ is centred near $\phi_{0}$. It is important to realise
that the estimate in Eq.~(\ref{opt_estimate}) is not optimal in an
absolute sense. Rather, it is the best estimate of $\phi$ given
that we have chosen to minimise the `cost function' $V^{H}(\phi)$.

%
%our approach is different to that of others
%

It is interesting to note that the approach to estimating
$\phi(t)$ outlined above differs from that in other work on phase
estimation \cite{wiseman95,wiseman97,wiseman98,berry02}. These
other papers generated estimates based on intuitive, partially
justified mathematical functions and, as a consequence, their
estimates were sometimes sub-optimal. In contrast, a number of
this paper's phase-estimation schemes use quantum parameter
estimation which leads to optimal estimates for $\phi(t)$ (at
least according to the cost or error function $V^{H}(\phi)$).

%
%how we use Bayes' rule in the cw scenario
%

To illustrate our method of obtaining ${\hat \phi}(t)$ via quantum
parameter estimation, we now demonstrate its application in the
case of measuring the signal beam via heterodyne detection. (Its
use in the other cases we consider is very similar.) For this type
of detection, Bayes' rule is
\begin{equation}
P(\phi|I_{\rm c}) = \frac{P(\phi) P(I_{\rm c}|\phi)}{P(I_{\rm
c})}.
\end{equation}
Replacing the normalisation constant $P(I_{\rm c})$ by $P(I_{\rm
c})_{|\alpha|=0}$ leads to the quasi-Bayes' rule
\begin{equation} \label{modified_bayes}
\tilde{P}(\phi|I_{\rm c}) =\frac{\tilde{P}(\phi) P(I_{\rm
c}|\phi)}{P(I_{\rm c})_{|\alpha|=0}}.
\end{equation}
Eq.~(\ref{hetero_measurement}) tells us that the real and
imaginary parts of $I_{\rm c}$ are Gaussian random variables with
variances of $1/(2dt)$ and, respectively, means of $|\alpha| \cos
\phi$ and $|\alpha| \sin \phi$. From this it follows that
\begin{eqnarray} \nonumber \label{expressn_one}
 \tilde{P}(I_{\rm c}|\phi) & = & \frac{dt}{\pi} \exp \left[ -dt \{ ({\rm
Re}(I_{\rm c})
 -|\alpha| \cos
\phi)^{2} \right. \\
& & \left.
 + ({\rm Im}(I_{\rm c}) - |\alpha| \sin \phi)^{2} \}/2
\right]
\end{eqnarray}
whilst
\begin{equation} \label{expressn_two}
P(I_{\rm c})_{|\alpha|=0}= \frac{dt}{\pi} \exp(-\frac{dt}{2}
I^{*}_{\rm c} I_{\rm c}).
\end{equation}
Substituting the expressions on the right-hand side of
Eqs~(\ref{expressn_one}) and (\ref{expressn_two}) into
Eq.~(\ref{modified_bayes}) and neglecting terms of order $dt^{2}$
or higher leads to the Zakai equation
\begin{equation} \label{het_zakai_eq}
d\tilde{P}(\phi) = |\alpha| (e^{i\phi}  I_{\rm c} + {\rm c.c.})
\tilde{P}(\phi)dt.
\end{equation}
Normalising $\tilde{P}$ via the known procedure detailed in
Appendix A, from Eq.~(\ref{het_zakai_eq}) we obtain the
Kushner-Stratonovitch (KS) equation \cite{mcgarty74}
\begin{eqnarray} \label{het_kse} \nonumber
d P (\phi) & = & |\alpha | \left[(e^{i\phi} -\bra e^{i\phi}
\ket_{P(\phi)}) P(\phi) \zeta(t)  + {\rm c.c.} \right]dt, \\
\end{eqnarray}
where $\zeta$ is complex Gaussian white noise ($\zeta=I_{\rm c} -
|\alpha| \bra  e^{i \phi} \ket_{P(\phi)}$ and is so-called
observation or measurement noise \cite{doherty00}). Incorporating
the effects of phase diffusion, we arrive at
\begin{eqnarray} \label{het_kse_full} \nonumber
d P(\phi) = \frac{\kappa}{2} \frac{\partial^{2} P(\phi) } {d
\phi^{2}}dt
\;\;\;\;\;\;\;\;\;\;\;\;\;\;\;\;\;\;\;\;\;\;\;\;\;\;\;\;\;\;\;\;\;\;\;\;\;\;\;\;\;\;\;\;\;&
& \\ \nonumber + |\alpha | \left[(e^{i\phi} -\bra e^{i\phi}
\ket_{P(\phi)})
P(\phi) \zeta(t) + {\rm c.c.} \right]dt. \\
\end{eqnarray}
Note that this equation has been previously derived, albeit for a
different (but related) physical system via a different method
\cite{wiseman93}. It is also interesting to realise that we could
have obtained Eq.~(\ref{het_kse_full}) via beginning with
Eq.~(\ref{bayes_rule}), substituting into it expressions for
$P(I_{\rm c}|\phi)$ and $P(I_{\rm c})$, and performing some
algebra whilst neglecting terms of order $dt^{2}$ or higher.
Though this method is conceptually simpler than the one we used,
it involves a more challenging calculation.
%Finally, a third way
%of arriving at Eq.~(\ref{het_kse_full}) would have been derive it
%from the system's stochastic master equation, as was done in
%\cite{wiseman93}.
To complete the process of determining ${\hat \phi}$, once we have
obtained Eq.~(\ref{het_kse_full}) we solve it and then use
$P(\phi)$ to calculate ${\hat \phi(t)}$ via
Eq.~(\ref{opt_estimate}).

%
%our approach is different to that of others
%

\section{Phase Estimation Schemes} \label{new_theory_sec}
In this paper we compare the accuracies of a number of
non-adaptive and adaptive phase estimation schemes for an EM beam.
Prior to doing so, however, we outline the schemes considered,
detailing non-adaptive and adaptive schemes in turn. These are
summarised in Table~\ref{table}.

\subsection{Non-adaptive schemes}

\subsubsection{Berry-Wiseman (BW) heterodyne-based}
In the {\em Berry-Wiseman (BW) heterodyne-based} phase-estimation
scheme \cite{berry02} the signal beam is measured via heterodyne
detection. The phase estimate at time $t$, ${\hat \phi}(t)$, is
then calculated from the measurement record up to $t$.
Specificially, it is
\begin{equation} \label{bw_het_est}
{\hat \phi}(t) = {\rm arg} \left( A_{t} \right),
\end{equation}
where $A_{t}$ can be written as
\begin{equation} \label{def_a}
A_{t} = \int_{u=-\infty}^{t} du e^{\chi(u - t)} I_{\rm c}(u),
\end{equation}
where $\chi$ is a scaling parameter. More specifically, $\chi$
scales the weight $\exp(-\chi (u-t))$ given to each current
$I_{u}$. Whilst this estimate may not seem intuitive, it was
chosen as an analogous estimate for the single-shot scenario was
known to be accurate \cite{wiseman97}. Moreover, Ref.
\cite{berry02} showed that, for large $N$, ${\rm arg}A_{t}$ was an
accurate estimate for a continuous EM beam when $\chi$ was set to
$\chi = 2 |\alpha| \sqrt{\kappa}$.

\subsubsection{Optimal heterodyne-based}
In this scheme, the signal beam is measured via heterodyne
detection and then, following  the calculation in
Section~\ref{theory_sec}, quantum parameter estimation is used to
obtain the KS equation Eq.~(\ref{het_kse}). This is then solved
and its solution used to obtain ${\hat \phi}(t)$ in accordance
with Eq.~(\ref{opt_estimate}).

\subsubsection{Canonical} \label{theory_canonical}
The {\em canonical} phase estimation scheme involves making a
canonical phase measurement \cite{leonhardt95} on the signal beam
at each instant in time and then taking ${\hat \phi}(t)$ to be its
outcome. Naively, it might be thought that this scheme would be
more accurate than any other as a canonical measurement, or so it
is thought, is the most accurate measurement of phase one can
make. Results in Section~\ref{results_sec} show, however, that
this is not the case (for reasons explained in
Section~\ref{discussion}).

\subsection{Adaptive schemes}

\subsubsection{Simple adaptive}
In the {\em simple adaptive} phase-estimation scheme
\cite{berry02} we measure the signal beam via adaptive homodyne
detection and then estimate $\phi(t)$ to be
\begin{equation} \label{simple_adapt}
{\hat \phi}(t) = {\rm arg} \left( A_{t} \right),
\end{equation}
where here
\begin{equation} \label{def_a_bar}
A_{t} = \int_{u=-\infty}^{t} du e^{\chi(u - t)} e^{i\Phi} I_{\rm
r}(u).
\end{equation}
We also adapt the homodyne measurement, setting the local
oscillator's phase to $\Phi(t)={\hat \phi}(t) + \pi/2$. From this
it follows \cite{berry02} that it is updated such that its rate of
change with time is
\begin{equation}
\frac{\partial \Phi}{\partial t} = {\sqrt \kappa} I_{\rm r}(t).
\end{equation}
This equation follows from letting $\chi =2|\alpha|\sqrt{\kappa}$
in Eq.~(\ref{def_a_bar}) which is known to be optimal for large
$N$ \cite{berry02}. One of the reasons the simple adaptive scheme
was considered in Ref. \cite{berry02} was that the fact that for
large $N$ it was known to be optimal. In Section~\ref{results_sec}
we show that it also performs well for small to moderates values
of $N$.

\subsubsection{Berry-Wiseman (BW) adaptive}
The {\em Berry-Wiseman (BW) adaptive} phase-estimation scheme
involves measuring the signal beam via adaptive homodyne
detection. The phase estimate at time $t$, ${\hat \phi}(t)$, is
then a function of two functionals of all measurement results up
to time $t$. Specifically, it is
\begin{equation} \label{bw_het_est_two}
{\hat \phi}(t) = {\rm arg} \left( A_{t} + \chi B_{t} A_{t}^{*}
\right),
\end{equation}
where $A_{t}$ is as defined in Eq.~(\ref{def_a}) and
%\begin{equation} \label{def_a}
%A_{t} = \int_{u=-\infty}^{t} du e^{\chi(u - t)}e^{i \Phi(u)}
%I_{\rm real}(u),
%\end{equation}
%where $\Phi(u)$ is the local oscillator phase at time $u$ and
$B_{t}$ is
\begin{equation}
B_{t} = \int_{u=-\infty}^{t} du e^{\chi(u - t)} e^{2i \Phi(u)}.
\end{equation}
As for the BW heterodyne-based scheme, this estimate was chosen as
an analogous estimate for the single-shot case was known to be
accurate \cite{wiseman97}. Furthermore, Ref. \cite{berry02} showed
that it was accurate, for large $N$, for $\chi = 2|\alpha|
\sqrt{\kappa}$

\subsubsection{Semi-optimal adaptive}
In the {\em semi-optimal adaptive} scheme for phase estimation, we
assume it is optimal to always measure the signal beam's phase
quadrature and thus, as in the other adaptive schemes, set
$\Phi(t) ={\hat \phi}(t) + \pi/2$. The reason we use the label
`semi-optimal adaptive' is that, whilst we use quantum parameter
estimation in determining ${\hat \phi}$, we are not certain that
it is always best to attempt to measure the phase quadrature.
Perhaps, one could obtain a more accurate estimate by occasionally
trying to measure the amplitude quadrature, for example.

\section{Results} \label{results_sec}

%
%V^H fluctuates & then settles down. We calculate V_steady_state
%

To compare the accuracies of the estimates introduced in
Section~\ref{new_theory_sec}, we now calculate their average
errors as measured by the Holevo variance $V^{H}$ of the
difference between the actual phase $\phi$ and our estimate ${\hat
\phi}$. Typically, this quantity fluctuates for some time before
settling down to a fixed steady-state value. Intuitively, this
occurs as a balance arises (on average) between the information we
gain about $\phi$ from a new photocurrent measurement and that we
lose due to $\phi$'s phase diffusion over the measurement's
duration.  We choose this steady-state value of $V^{H}(\phi-{\hat
\phi})$, denoted by $V^{H}_{SS}$, as our measure of the efficacy
of our phase-estimation schemes and hence numerically determine it
for all of them for a range of $N$ values. We also obtain analytic
expressions for it for some schemes for both large and small
values of $N$.

%
%how we actually calculate V^H_SS
%

From the definition of the Holevo variance in
Eq.~(\ref{holevo_define}), $V^{H}(\phi -{\hat \phi})$ is given by
\begin{eqnarray}
V^{H} (\phi -{\hat \phi}) & = & | \bra e^{i(\phi -{\hat \phi})}
\ket_{\xi,I} |^{-2}-1,
\end{eqnarray}
where the average $\bra \ldots \ket_{\xi,I}$ is a stochastic
average over $\xi$ and $I$. To calculate this quantity for our
three estimates generated via parameter estimation, we first
%he Holevo variance $V^{H}(\phi-{\hat \phi})$ was calculated via
use the fact that
\begin{equation} \label{identity}
\bra e^{i(\phi -{\hat \phi})} \ket_{\xi,I} = \bra | \bra e^{i
\phi} \ket_{P(\phi)} | \ket_{I}
\end{equation}
%where $\bra \ldots \ket_{\zeta}$ is a stochastic average over
%$\zeta$,
to express $V^{H}(\phi-{\hat \phi})$ as
\begin{eqnarray} \label{step_one_holevo}
V^{H}(\phi -{\hat \phi}) & = & \bra  | \bra e^{i\phi}
\ket_{P(\phi)} | \ket_{I} ^{-2}-1.
\end{eqnarray}
A demonstration of Eq.~(\ref{identity}) is given in Appendix B.
%Its utility lies in the fact that it allows us to ignore the true
%phase $\phi$ when calculating $V^{H}_{SS}$.
%right-hand side is easier to calculate than its left-hand side by
%virtue of being independent of the actual phase $\phi$.
%Employing this identity,
%$V^{H}_{SS}$ becomes
%\begin{eqnarray}
%V^{H}_{SS} & = & | \bra e^{i(\phi -{\hat \phi})} \ket_{\xi,\zeta}
%|^{-2}-1 \\
%& = & \bra | \bra e^{i \phi} \ket_{P(\phi)} | \ket_{\zeta}.
%\end{eqnarray}
After arriving at Eq.~(\ref{step_one_holevo}), we then use the
ergodic theorem within this equation to replace the ensemble
average $\bra | \bra e^{i \phi} \ket_{P(\phi)} | \ket_{I}$ {\em in
the steady-state} by the temporal average
\begin{equation} \label{temporal_average}
\frac{1}{t_{f} - t_{0}^{SS}} \int_{t=t_0^{SS}}^{t_{f}} dt | \bra
e^{i{\hat \phi}} (t) \ket_{P(\phi)} |,
\end{equation}
where $t_{0}^{SS}$ is the time at which the steady-state regime
begins and $t_{f}$ is the final time we consider ($t_{f} \gg
t_{0}^{SS}$).
%we calculated with the ensemble average $\bra
%\exp(i(\phi(t)-{\hat \phi}(t))) \ket_{SS}$
This allows us to determine $V^{H}_{SS}$ through simulating just a
single stochastic trajectory.
%Finally, $V^{H}_{SS}$ is calculated
%via numerically determining expression~(\ref{temporal_average})
%and then substituting the result into Eq.~(\ref{step_one_holevo}).
%Note that in a communication scenario this quantity will tell us
%something about the amount of information we can transmit in the
%steady-state using a particular phase encoding and a particular
%phase-estimation scheme. Given these considerations, we
%numerically determine $V^{H}(\phi-{\hat \phi})$ in the
%steady-state, which we denote by $V_{SS}^{H}$, for the schemes in
%Section~\ref{new_theory_sec}.

%
%\hat \phi has initial transience & then locks onto \phi
%

\begin{figure}
\center{\epsfig{figure=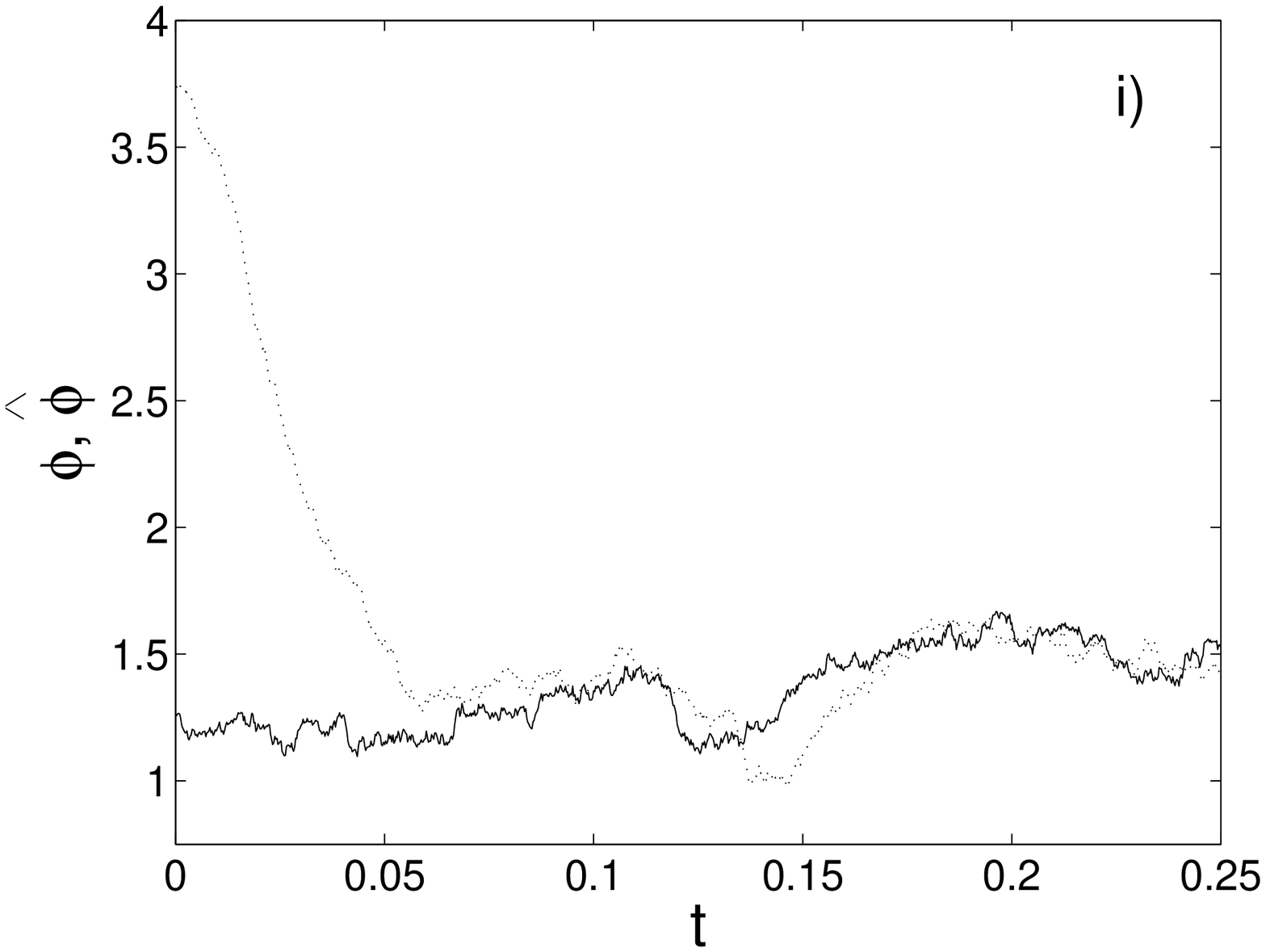,width=80mm}}
\center{\epsfig{figure=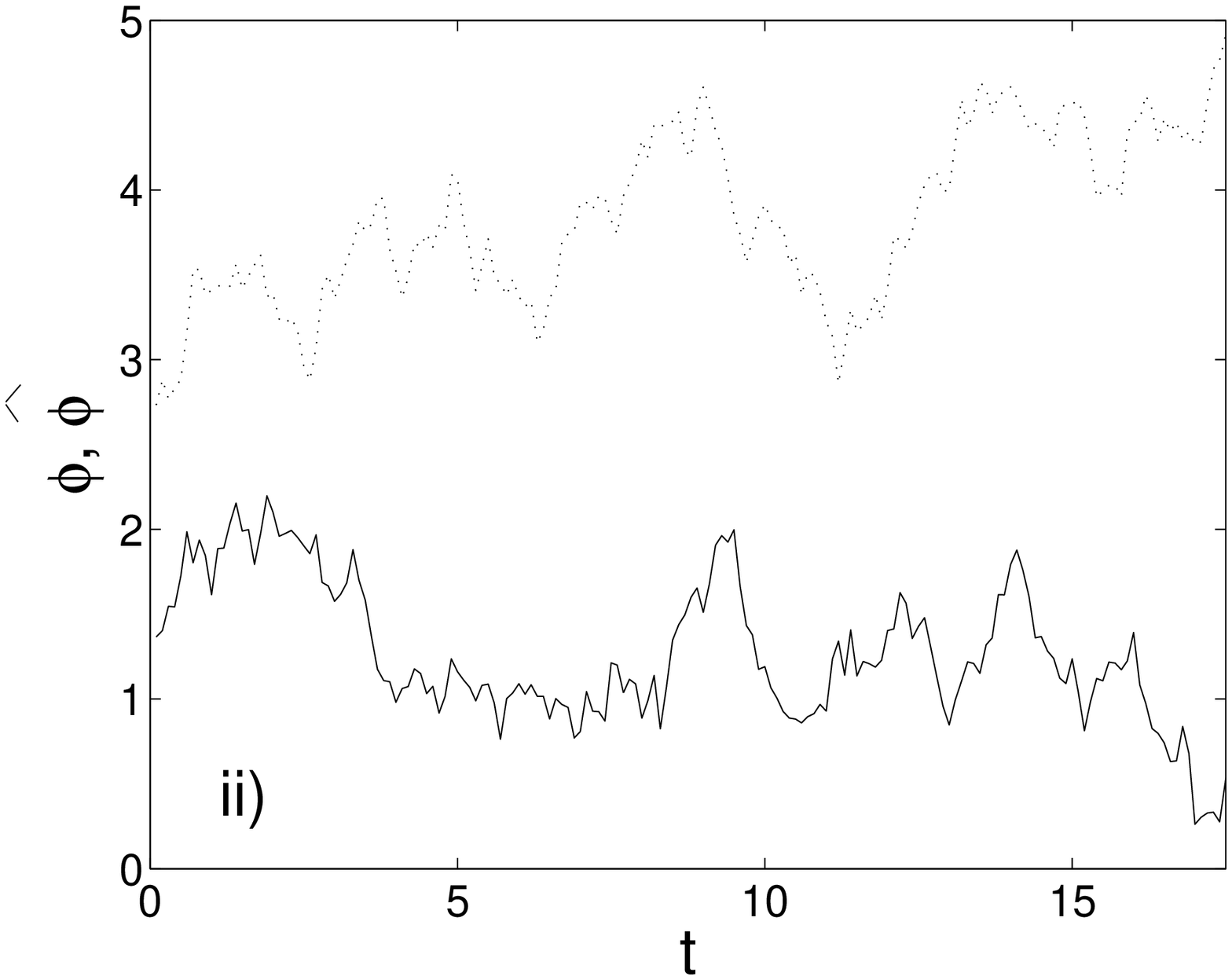,width=80mm}} \caption{Graphs
showing typical variations of the actual phase $\phi$ (solid line)
and our estimate ${\hat \phi}$ (dotted line) versus time $t$
scaled by $\kappa$ for the simple-adaptive phase-estimation scheme
for: i) a large photon flux ($N=1000$) and ii) a small one
($N=0.1$). In i), ${\hat \phi}$ initially `homes in' on $\phi$,
before locking onto it. In ii), the low photon flux means we gain
so little information from our measurements that ${\hat \phi}$
never locks onto $\phi$. Both $\phi$ and ${\hat \phi}$ are
dimensionless, as is $t$.} \label{locking}
\end{figure}

Upon calculating $V^{H}_{SS}$, a number of trends are apparent.
%In the adaptive schemes we consider the fact that $V^{H}(\phi-{\hat
%\phi})$ settles down to the steady-state value $V_{SS}^{H}$ can be
%viewed as arising from the following behaviour of ${\hat \phi}$:
The first of these concerns the proximity of ${\hat \phi}$ to
$\phi$ in the simple adaptive scheme. For large $N$,
%the estimate
%${\hat \phi}$ typically 'homes in' quickly on the true phase
%$\phi$ as illustrated for the simple adaptive scheme in
%Fig.~\ref{locking} i). The
the initial estimate ${\hat \phi}(t=0)$ for this scheme is usually
some distance from the actual phase $\phi(t=0)$. Then, as we gain
more and more information via measurement and post-processing,
${\hat \phi}$ `homes in' on $\phi$ during an initial period of
transience.
%and then homes in on it during an initial period of transience.
After this it `locks onto' $\phi$, staying close to $\phi$ as it
continues to fluctuate a little.
%There still are,
%however, some residual fluctuations in the difference between
%${\hat \phi}$ and $\phi$.
This pattern of behaviour is illustrated in Fig.~\ref{locking} i).
It is anticipated that all the schemes considered behave
similarly, though we did not explicitly verify this. For small
values of $N$, ${\hat \phi}$ never locks onto $\phi$ but instead
continues to fluctuates in its vicinity with a magnitude that
increases with decreasing $N$, as highlighted in
Fig.~\ref{locking} ii).
%In both cases, the difference between
%$\phi$ and ${\hat \phi}$ tends to settle down to a fixed value (on
%average) after some time. In turn, this leads to $V^{H}(\phi-{\hat
%\phi})$ also settle down to a steady-state value.
\begin{figure}
\center{\epsfig{figure=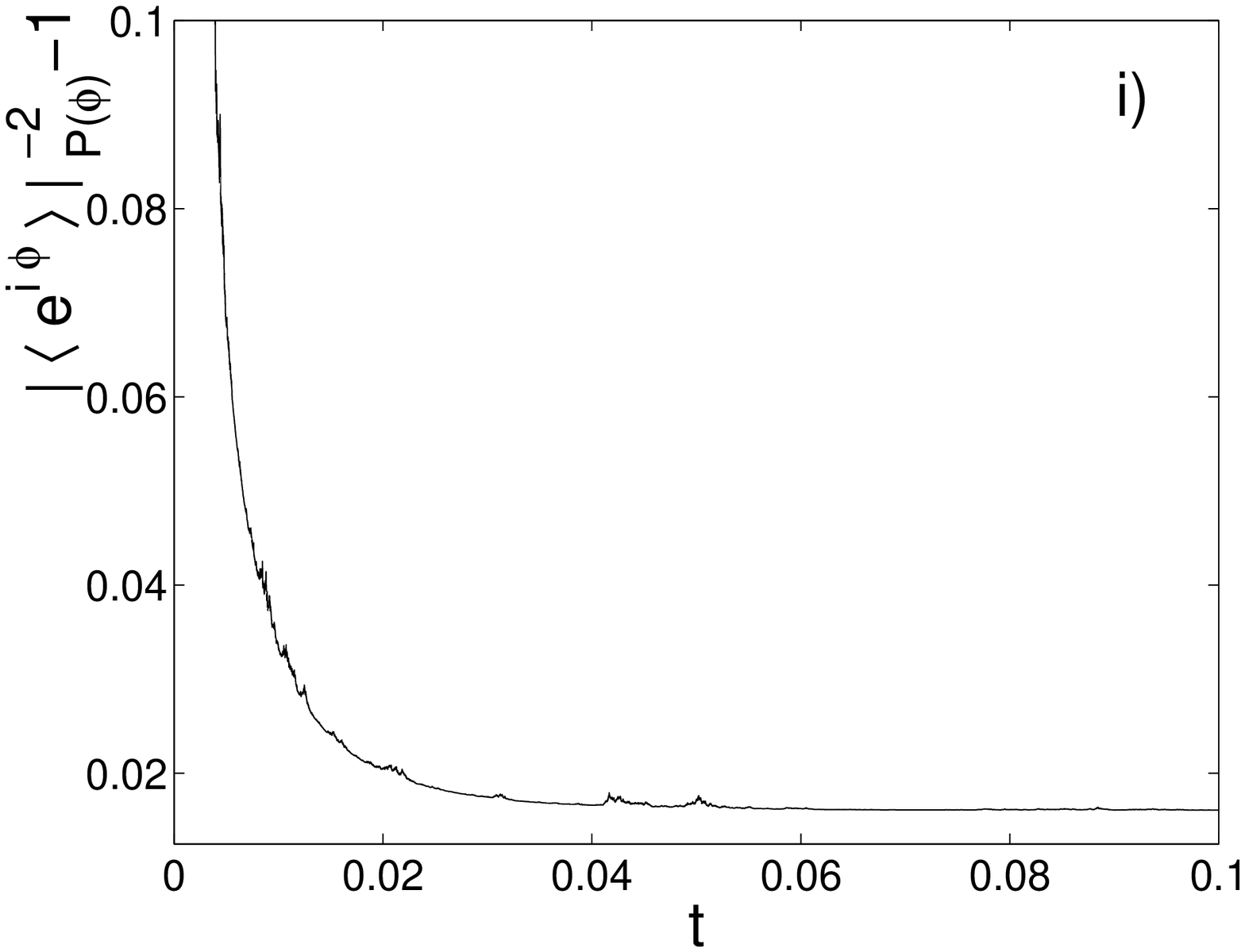,width=80mm}}
\center{\epsfig{figure=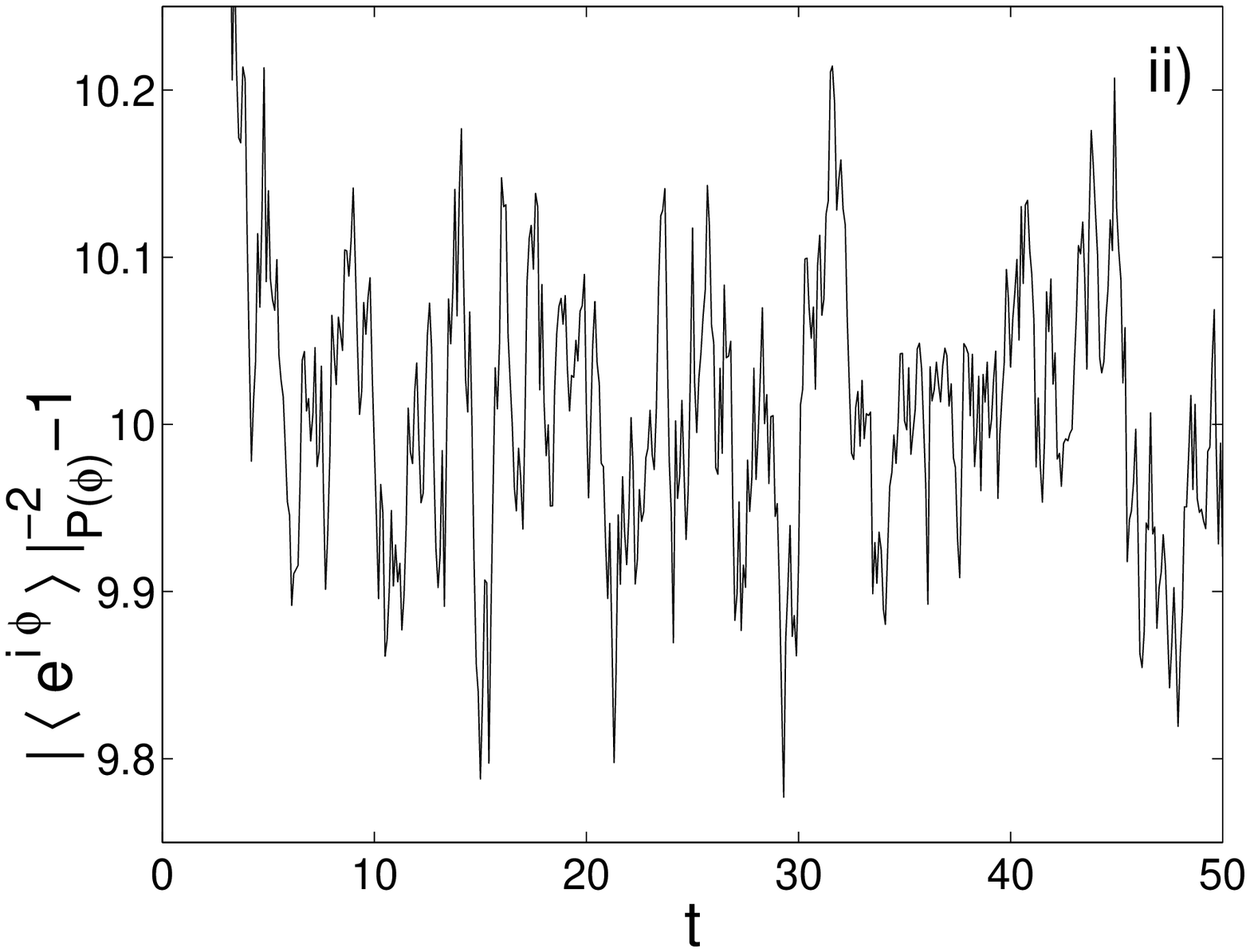,width=80mm}} \caption{Graphs
showing typical variations of our lack of confidence in ${\hat
\phi}$ as measured by $| \bra e^{i\phi} \ket_{P(\phi)} |^{-2}-1$
(dimensionless) versus time $t$ (dimensionless) scaled by $\kappa$
for the semi-optimal adaptive phase-estimation scheme for: i) a
large photon flux ($N=1000$) and ii) a small one ($N=0.1$).}
\label{confidence}
\end{figure}

A second trend in our results concerns the size of the interval
within which we are fairly certain that $\phi$ lies at any moment
in time. This is measured by the Holevo variance $| \bra e^{i
\phi} \ket_{P(\phi)} |^{-2} - 1$ which can be thought of as a
measure of our lack of confidence in ${\hat \phi}$. For large $N$,
this quantity, at least for the schemes based on parameter
estimation, only fluctuates over time by a small amount once the
initial transience ends. This behaviour is illustrated in
Fig.~(\ref{confidence}) i). It can be explained by realising that
when $N$ is large we are in a linear regime in the sense that the
measured photocurrent, $I_{\rm r}$ or $I_{\rm c}$, is a {\em
linear} function of the actual phase $\phi$. For instance, for
homodyne detection we have $I_{\rm r}dt= 2\eta |\alpha| (\phi -
{\hat \phi})  + \sqrt{\eta} dW$. It is a characteristic trait of
such linear systems that our level of confidence (and hence also
our lack of confidence) in any estimate of a system parameter is
constant in the steady state \cite{mcgarty74}. For small $N$,
however, $| \bra e^{i \phi} \ket_{P(\phi)}|^{-2}-1$ fluctuates
appreciably for all $t$ (for the schemes based on parameter
estimation), as shown in Fig.~(\ref{confidence}) ii).

\subsection{Non-adaptive schemes}

\subsubsection{Berry-Wiseman heterodyne-based}

Previous work \cite{berry02} has calculated $V^{H}_{SS}$ for the
BW heterodyne-based scheme for a range of $N$ values. These
results are plotted in Fig.~\ref{graph_one}. For large $N$, it is
known \cite{berry02} that the scheme has a steady-state error of
$V^{H}_{SS} \simeq 1/\sqrt{2 N}$.

\subsubsection{Optimal heterodyne-based} \label{opt_het_subsub}
\begin{figure}
\center{\epsfig{figure=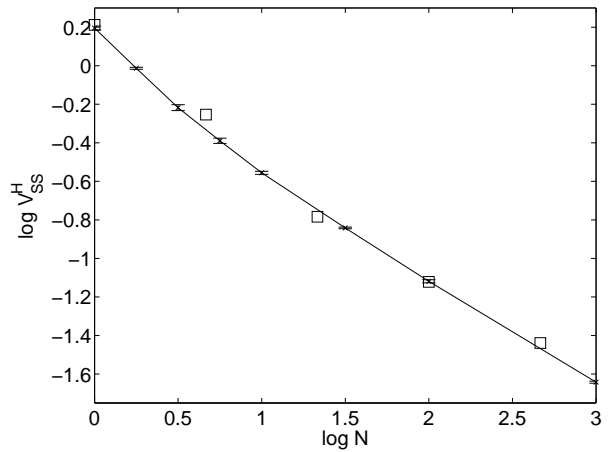,width=80mm}} \caption{Log-log
plot of the steady-state Holevo variance $ V^{H}_{SS}$ versus
photon flux $N$ for the BW heterodyne-based non-adaptive (squares)
and optimal heterodyne non-adaptive (solid line) phase estimation
schemes. Both $V^{H}_{SS}$ and $N$ are dimensionless.}
\label{graph_one}
\end{figure}

For the optimal heterodyne-based phase-estimation scheme,
$V^{H}_{SS}$ was calculated
%via calculating $\bra | \bra e^{i \phi}
%\ket_{P(\phi)} | \ket_{\zeta}$ and then using the fact that it is
%equal to $\bra e^{i(\phi -{\hat \phi})} \ket_{\xi,\zeta}$. In
%particular, this Holevo variance was found
by determining the temporal average in expression
\ref{temporal_average} and then using Eq.~(\ref{step_one_holevo})
to find $V^{H}_{SS}$. This was done by, first, expressing
$P(\phi)$ in Eq.~(\ref{het_kse_full}) as the following discrete
Fourier series:
\begin{equation} \label{fourier_decomp}
P(\phi) = \sum_{j=-\infty}^{\infty} b_{j} \exp(ij \phi),
\end{equation}
where $b_{j} \in {\mathbb C}$ and $b_{-j}=b_{j}^{*}$. Next, the
resulting equation was transformed into Fourier space to produce
the following coupled differential equations:
\begin{equation} \label{fourier_de}
{\dot b}_{j} = -\frac{\kappa j^{2} b_{j}}{2} + |\alpha|  \zeta
b_{j-1} + |\alpha|  \zeta^{*} b_{j+1} -4 \pi b_{j} |\alpha| {\rm
Re} \left(
 \zeta^{*} b_{1} \right).
\end{equation}
These were then numerically solved by only considering $b_{j}$'s
for which $|j|$ was less than some finite bound that increased
with $N$. Next, $\bra e^{i\phi} \ket_{P(\phi)}(t)$ was determined
by exploiting the fact that it is a function of just one Fourier
coefficient ($|b_{1}|$). Finally, we averaged over numerous
steady-state values of $\bra e^{i\phi} \ket_{P(\phi)}(t)$ to
obtain expression~(\ref{temporal_average}) and thus $V^{H}_{SS}$.
The results generated are plotted in Fig.~\ref{graph_one}.
Analytic results were also found for large and small $N$ which are
$V^{H}_{SS} \simeq 4/(\pi N)$ (small $N$) and $V^{H}_{SS} \simeq
1/\sqrt{2 N}$ (large $N$).

%
%summary of the small-N method
%

%The small-$N$ result for $V^{H}_{SS}$ was obtained from
%Eq.~(\ref{het_kse_full}) by first transforming it into a discrete
%Fourier space based on the decomposition $P(\phi) = \sum_{j} b_{j}
%e^{i j \phi}$, where the $b_{j}$ are Fourier coefficients. The
%resulting coupled differential equations with respect to time for
%the Fourier coefficients of $P(\phi)$ then simplified for the
%following reason: When $N \ll 1$ the photon flux is small and so
%measurements tell us very little about $\phi$, meaning that
%$P(\phi)$ is broad. Conversely, the Fourier transform of $P(\phi)$
%is narrow and so only a small number of Fourier coefficients are
%appreciable. This allowed us to simplify the coupled differential
%equations for the Fourier coefficients and then to solve them.
%%When $N \ll 1$ only a few coefficients contribute appreciably to
%%$P(\phi)$'s solution and thus we were able to simplify the
%%differential equations for the appreciable Fourier coefficients
%%and then solve them.
%Having done this, the resulting solution for one of the components
%was then used to $P(\phi)$ to calculate $V^{H}_{SS}$ via the
%formula $V^{H}_{SS} = \bra | \bra e^{i \phi}\ket_{P(\phi)} |
%\ket_{\zeta}^{-2} - 1$.

%
%details of the small-N method
%

Our analytic result for $V^{H}_{SS}$ for small $N$ was obtaining
by first realising that when $N \ll 1$ heterodyne measurements on
the signal beam yield little information about $\phi$ and thus
$P(\phi)$ is broad. This means that, in contrast, $P(\phi)$'s
Fourier transform is narrow and, more specifically, that the
following relations hold (on average) $|b_{0}| \gg |b_{1} | \gg
|b_{2}| \ldots$. Because of this, we can neglect Fourier
coefficients for which $|j| > 1$ in Eq.~(\ref{fourier_de}). Upon
doing this, and also neglecting terms containing $|b_{1}|^{2}$ (as
$|b_{1}|^{2} \ll |b_{1}|$), we are left with just the following
equation for $b_{1}$
\begin{equation} \label{b_de}
\frac{\partial b_{1}}{\partial t} = - \frac{\kappa b_{1}}{2}
-\frac{|\alpha| \zeta}{2 \pi}.
\end{equation}
Note that $b_{0}(t)=1/(2\pi)$ (as can be determined from the
normalisation condition $\int_{\phi=\phi_{0}}^{2\pi + \phi_{0}}
P(\phi)d\phi=1$). Solving Eq.~(\ref{b_de}) we find that in the
steady-state $b_{1}$ is a complex Gaussian random variable with
mean zero and a variance of $N/(8 \pi^{2})$ in both its real and
imaginary parts.

To calculate $V_{SS}^{H}$ from $b_{1}$ we first note that
\begin{equation}
|\bra e^{i\phi} \ket_{P(\phi)}| =
|\int_{\phi=\phi_{0}}^{\phi_{0}+2 \pi} d\phi e^{i\phi} P(\phi) |.
\end{equation}
Substituting $P(\phi) = \sum_{j=-\infty}^{\infty} b_{j} \exp(ij
\phi)$ into the right-hand side of this equation yields
\begin{equation}
|\bra  e^{i\phi}  \ket_{P(\phi)}| =  2 \pi | b_{1} |.
\end{equation}
From this it follows that the equation
\begin{equation}
V^{H}_{SS} = (\bra | \bra  e^{i\phi} \ket_{P(\phi)}| \ket_{I}^{-2}
- 1,
\end{equation}
simplifies to
\begin{equation}
V^{H}_{SS} = \left( 2\pi \bra |b_{1}| \ket_{\zeta} \right)^{-2} -
1.
\end{equation}
%from which it follows that to calculate $V^{H}_{SS}$ the only
%moment we need to determine is $\bra |b_{1}| \ket$.
Given that $\bra |b_{1}| \ket_{\zeta} \simeq
\sqrt{N}/(4\sqrt{\pi})$ we obtain
%\begin{equation}
%V^{H}_{SS} = \frac{}{\left( 2\pi \bra |b_{1}| \ket \right)^{2}} -
%1
%\end{equation}
$V^{H}_{SS} \simeq 4/(\pi N) -1$. Neglecting the second term (as
this produces a more accurate approximation) yields
\begin{equation} \label{small_N_het}
V^{H}_{SS} \simeq 4/(\pi N).
\end{equation}

%
%getting large-N result
%

The large-$N$ approximation for $V^{H}_{SS}$ for the optimal
heterodyne scheme was obtained by replacing the exponents in
Eq.~(\ref{het_kse_full}) by a linear approximation and then
assuming that $P(\phi)$ was Gaussian. Differential equations with
respect to time for the mean and variance of this Gaussian were
then constructed and solved to obtain the standard variance of
${\hat \phi}$ in the steady state which, for large $N$, is
approximately equal to $V^{H}_{SS}$.

%Given the fact that
%\begin{equation}
%e^{i {\hat \phi}} = \frac{\bra e^{i\phi} \ket}{|\bra e^{i\phi}
%\ket|},
%\end{equation}
%it follows that
The expression $\left( e^{i\phi} - \bra e^{i\phi} \ket_{P(\phi)}
\right) \zeta$ in Eq.~(\ref{het_kse_full}) can be re-expressed as
\begin{equation} \label{term}
(e^{i(\phi - {\hat \phi}) } - \bra e^{i(\phi - {\hat \phi})
}\ket_{P(\phi)}) e^{i {\hat \phi}} \zeta.
\end{equation}
When $N \gg 1$, the large photon fluxes present in the signal beam
mean that our measurements yield a great deal of information about
$\phi$ and hence that ${\hat \phi}$ is a highly accurate estimate.
As a result, $e^{i(\phi - {\hat \phi})} \simeq 1$ and thus we can
linearise expression~(\ref{term}) as follows
\begin{equation}
(e^{i(\phi - {\hat \phi}) } - \bra e^{i(\phi - {\hat \phi})
}\ket_{P(\phi)}) e^{i {\hat \phi}} \zeta \simeq i( \phi - \bra
\phi \ket_{P(\phi)}) e^{i {\hat \phi}} \zeta.
\end{equation}
The expression $e^{i {\hat \phi}} \zeta$ behaves as complex
Gaussian white noise and hence we denote it as $\zeta'$.
Substituting the above results into Eq.~(\ref{het_kse_full}), we
obtain
%\begin{equation}
%dP = \frac{\kappa}{2}
%\end{equation}
%\begin{eqnarray} \label{linearised}
%d P(\phi) & = & \frac{\kappa}{2} \frac{\partial^{2} P } {d
%\phi^{2}}dt
%\\ \nonumber & - & \sqrt{2} |\alpha | \left( i(\phi - \bra
%\phi \ket )\zeta_{\rm real}' \right) dt,
%\end{eqnarray}
\begin{eqnarray} \label{linearised} \nonumber
d P(\phi) & = & \frac{\kappa}{2} \frac{\partial^{2} P } {d
\phi^{2}}dt -2 |\alpha | \left( i(\phi - \bra \phi \ket_{P(\phi)}
) {\rm Re}(\zeta') \right) dt. \\
\end{eqnarray}
To solve this equation, we assume that $P(\phi)$ is Gaussian and
thus that it can be expressed as
%is characterised
%by its mean $\mu_{P$ and its variance $\sigma^{2}_{P}$.
%Substituting the solution
\begin{equation}
P(\phi)= \frac{ \exp\left[-(\phi-\mu_{P})^{2}/(2
\sigma^{2}_{P})\right] }{\sqrt{2 \pi} \sigma_{P}},
\end{equation}
where $\mu_{P}$ and $\sigma_{P}^{2}$ are, respectively, $P$'s mean
and variance.
%Substituting in this solution for $P$ into
%Eq.~(\ref{linearised}), we obtain differential equations for
Generating differential equations for $\mu_{P}$ and $\sigma_{P}$,
we obtain,
\begin{eqnarray} \nonumber
d \sigma_{P}^{2} & = & d \bra \phi^{2} \ket_{P(\phi)} - d(\bra
\phi\ket^{2}_{P(\phi)} ) \\ \nonumber & = & d \bra \phi^{2}
\ket_{P(\phi)} - 2 \bra \phi \ket_{P(\phi)}
d \bra \phi \ket_{P(\phi)} - (d \bra \phi \ket_{P(\phi)})^{2} \\
\end{eqnarray}
and
\begin{equation} \label{average}
d\mu_{P} = -2 |\alpha| \sigma_{P}^{2} {\rm Re}(\zeta') dt.
\end{equation}
Solving these yields
\begin{equation}
\sigma^{2}_{P}(t) =  \frac{1}{\sqrt{2N}} \frac{ \exp(2\sqrt{2}
|\alpha|^{2} t/\sqrt{N}) + 1}{\exp(2 \sqrt{2}
 |\alpha|^{2} t/\sqrt{N}) -1}.
\end{equation}
In the limit of $t \rightarrow \infty$ this reduces to
\begin{equation}
\sigma^{2}_{P\:SS} \simeq  V^{H}_{SS} \simeq \frac{1}{\sqrt{2 N}}.
\end{equation}
Interestingly, this result is the same as that obtained in
\cite{berry02}. This shows that the BW heterodyne-based scheme,
which was designed for large $N$, is indeed optimal in this
regime.

\subsubsection{Canonical}
For the canonical phase estimation scheme, ${\hat \phi}(t)$ was
calculated via quantum parameter estimation using the method in
Subsection~\ref{parameter_estimation_theory}. For this scheme,
Bayes' rule is
\begin{equation} \label{bayes_canonical}
P(\phi|\theta) = \frac{P(\phi) P(\theta|\phi)}{P(\theta)},
\end{equation}
where $\theta$ is the measured phase.
%Replacing the
%normalisation constant $P(\theta_{\rm meas})$ by $P(\theta_{\rm
%meas})_{|\alpha|=0}$ yields
%\begin{equation} \label{semi_modified_bayes_canonical}
%\tilde{P}(\phi|\theta_{\rm meas})
%=\frac{\tilde{P}(\phi) P(\theta_{\rm
%meas}|\phi)}{P(\theta_{\rm meas})_{|\alpha|=0}}.
%\end{equation}
%Physically, $P(\theta_{\rm meas})_{|\alpha|=0}$ is the probability
%of measuring $\theta_{\rm meas}$ given that the amplitude of the
%field $|\alpha|$ is zero. In this scenario, the field's phase is
%undefined and thus all possible phase-measurement outcomes occur
%with probability $1/(\2\pi)$ and thus $P(\theta_{\rm
%meas})_{|\alpha|=0}=1/(2\pi)$.
%Eq.~(\ref{homo_measurement}) tells us that $I_{\rm r}$ is a
%Gaussian random variable with variance $1/(dt)$ and mean
%$2|\alpha| \cos \phi$.
As a canonical phase measurement is a projective measurement of
the Pegg-Barnett phase observable \cite{pegg88}, the probability
of it yielding the result $\theta$ is $(2\pi)^{-1}$ times the
square of the norm of the measured state's projection onto the
(unnormalised) phase eigenstate $| \theta \ket=
\sum_{n=0}^{\infty} e^{i n \theta} | n \ket$. Thus, for the
coherent states we consider, to first order in $\sqrt{dt}$,
\begin{eqnarray} \label{phase_measurement_overlap} \nonumber
P(\theta |\phi) & = & \frac{1}{2\pi} | \bra \alpha \sqrt{dt} | \theta \ket|^{2} \\
& = & \frac{1}{2\pi} \left( 1 + 2 |\alpha| \sqrt{dt} \cos(\theta -
\phi) \right)
\end{eqnarray}
and thus
%it
%follows that
%\begin{eqnarray} \nonumber
%\tilde{P}(\theta_{\rm meas} | \phi) &  = &  P(\phi) \left(
%1+ 2|\alpha| \sqrt{dt} \cos(\phi-\theta) \right).
%%\exp \left[ -dt \left( ({\rm Re}(I_{\rm r})
% -|\alpha| \cos
%\phi-\Phi)^{2} \right. \right. & & \\
%\left. \left. \right) \right] & &
%\end{eqnarray} and
\begin{equation} \label{phase_normalisation}
P(\theta)_{|\alpha|=0}= (2\pi)^{-1}.
\end{equation}
Substituting the expressions on the right-hand sides of
Eqs~(\ref{phase_measurement_overlap}) and
(\ref{phase_normalisation}) into Eq.~(\ref{bayes_canonical}) leads
to the following Zakai equation
\begin{equation} \label{canonical_zakai_eq}
d\tilde{P}(\phi) = \frac{|\alpha|}{\sqrt{dt}} (e^{i(\phi-\theta)}
+ {\rm c.c.}) \tilde{P}(\phi) dt.
\end{equation}
Using the known correspondence detailed in Appendix A, this, in
turn, leads to the KS equation
\begin{eqnarray} \label{kse_canonical} \nonumber
d P(\phi) = 2 \times  \;\;\;\;\;\;\;\;\;\;\;\;\;\;\;\;\;\;\;\;\;\;
\;\;\;\;\;\;\;\;\;\;\;\;\;\;\;\;\;\;\;\;\;\;\;\;\;\;\;\;\;\;\;\;\;\;\;\;\;\;\;\;\; & & \\
\nonumber \;\;\;\;\;\;\;\;\;\;\;\;\;\;\;{\rm Re} \left( |\alpha |
\left[(e^{i(\phi)} -\bra e^{i \phi} \ket_{P(\phi)})
\frac{e^{-i\theta}} {\sqrt{dt}}
P(\phi) \right]dt \right). & & \\
\end{eqnarray}
%where $\zeta'$ is real Gaussian white noise ($\zeta'=I_{\rm r}
%- 2 |\alpha| \bra cos(\phi-\Phi) \ket$).
Letting $e^{-i\theta}/\sqrt{dt} = f$, we find that
%$e^{-i\theta} \bra
%e^{i \phi}\ket/(\sqrt{dt} |\bra e^{i \phi}\ket|)$
$\bra f \ket = \bra f^{2} \ket =0$ (at least when we average over
any finite time interval) and $\bra f f^{*} \ket=1/dt$ from which
it follows that $f$ is complex Gaussian white noise. Given this,
Eq.~(\ref{kse_canonical}) reduces to Eq.~(\ref{het_kse}), the KS
equation obtained for the optimal heterodyne-based
phase-estimation scheme. As a result, the canonical scheme shares
the same accuracy as this other scheme and so shares the same
results for $V^{H}_{SS}$. This surprising result is explained in
Section~\ref{discussion}.

\subsubsection{Comparison}

%
%opt het. is slightly better than BW het.
%
As can be seen from Fig.~\ref{graph_one}, when $N \alt 10$, the
optimal heterodyne-based phase-estimation scheme is slightly more
accurate than the BW heterodyne-based one. For larger values of
$N$, however, we see that both schemes seem to be equally
accurate. (At approximately $N=10^{1.25}$, the BW heterodyne-based
scheme appears to be more accurate, but this is due to numerical
errors, primarily in the BW heterodyne-based result.) The first of
these features illustrates that while the BW heterodyne-based
scheme is close to optimal for $N \alt 10$, $\phi$ can be
estimated more accurately using parameter estimation in this
regime. The latter fact is particularly significant as this regime
is the one in which an experimental realisation could most readily
be performed, as discussed in more detail in
Sect.~\ref{discussion}. The second feature highlights that the BW
heterodyne-based scheme is optimal for $N \gtrsim 10$ which is
unsurprising as it was designed for large $N$ \cite{berry02}.

\subsection{Adaptive schemes}

\subsubsection{Simple adaptive}
For the simple adaptive phase-estimation scheme, the Holevo
variance in the steady-state was calculated by simulating the
evolution of $\phi(t)$ via solving Eq.~(\ref{true_phase}) and also
simulating the measurement outcomes on the beam using
Eq.~(\ref{homo_measurement}) to obtain a numerical expression for
$I_{\rm r}(t)$ for a range of times. This allowed us to update
${\hat \phi}$ via
%This current was used to update
%$\Phi(t)$ in accordance with
\begin{equation} \label{phi_update}
\frac{\partial {\hat \phi}}{\partial t} = {\sqrt \kappa} I_{\rm
r}(t)
\end{equation}
and thus to determine $\phi(t) - {\hat \phi}(t)$, again for a
range of times. The local-oscillator phase $\Phi(t)$ was then set
to $\Phi(t) = {\hat \phi}(t) + \pi/2$.
%and also to update ${\hat \phi}(t)$
%which shares the same evolution equation as $\Phi$.
The steady-state Holevo variance $V^{H}_{SS}$ was calculated from
the difference $\phi(t)-{\hat \phi}(t)$.
%In doing this, we invoked
%the ergodic theorem which allowed us to equate the temporal
%average $1/(t_{f} - t_{0}^{ss}) \int_{t=t_0^{ss}}^{\infty} dt
%\exp(i(\phi-{\hat \phi}))$, where $t_{0}^{ss}$ is the time at
%which the steady-state regime begins and $t_{f}$ is the final time
%we consider ($t_{f} \ge t_{0}^{ss}$), we calculated with the
%ensemble average $\bra \exp(i(\phi(t)-{\hat \phi}(t))) \ket_{SS}$,
%where $\bra \ldots \ket_{SS}$ denotes a steady-state average.

\subsubsection{Berry-Wiseman adaptive}

For the BW adaptive scheme, Ref. \cite{berry02} determined
$V^{H}_{SS}$ as a function of $N$ and these results are shown in
Fig.~\ref{graph_two}.

\subsubsection{Semi-optimal adaptive}
We derived ${\hat \phi}$ for the semi-optimal adaptive scheme via
quantum parameter estimation in the same manner as for the optimal
heterodyne and canonical schemes. For this scheme, Bayes' rule is
\begin{equation}
P(\phi|I_{\rm r}) = \frac{P(\phi) P(I_{\rm r}|\phi)}{P(I_{\rm
r})}.
\end{equation}
Replacing the normalisation constant $P(I_{\rm r})$ by $P(I_{\rm
r})_{|\alpha|=0}$ yields the quasi-Bayes rule
\begin{equation} \label{semi_modified_bayes}
\tilde{P}(\phi|I_{\rm r}) =\frac{\tilde{P}(\phi) P(I_{\rm
r}|\phi)}{P(I_{\rm r})_{|\alpha|=0}}.
\end{equation}
From Eq.~(\ref{homo_measurement}) we know that $I_{\rm r}$ is a
Gaussian random variable with variance $1/(dt)$ and mean
$2|\alpha| \cos (\phi - \Phi)$ from which it follows that (for
$\eta=1$)
\begin{eqnarray} \nonumber
P(\phi|I_{\rm r}) & = & \frac{dt}{\pi} \exp \left[ -dt \{ I_{\rm
r}
 -2 |\alpha| \cos
(\phi-\Phi) \}^{2} \right] \\
\end{eqnarray}
and
\begin{equation}
P(I_{\rm r})_{|\alpha|=0}= \frac{dt}{\pi} \exp(-dt I_{\rm r}^{2}).
\end{equation}
Substituting these two results into
Eq.~(\ref{semi_modified_bayes}), we obtain the following Zakai
equation
\begin{equation} \label{semi_zakai_eq}
d\tilde{P}(\phi) = |\alpha| (e^{i(\phi-\Phi)}  I_{\rm r} + {\rm
c.c.}) \tilde{P}(\phi)dt.
\end{equation}
Using the known correspondence detailed in Appendix A and
including the effects of phase diffusion,
Eq.~(\ref{semi_zakai_eq}) leads to the KS equation
\begin{eqnarray} \label{kse_semi} \nonumber
d P(\phi) & = & \frac{\kappa}{2} \frac{\partial^{2} P(\phi) } {d
\phi^{2}}dt + \\ \nonumber & & |\alpha | \left[(e^{i(\phi-\Phi)}
-\bra e^{i(\phi-\Phi)} \ket_{P(\phi)}) P(\phi) \zeta'(t) + {\rm
c.c.} \right]dt,
\\
\end{eqnarray}
where $\zeta'$ is {\em real} Gaussian white noise given by
$\zeta'=I_{\rm r} - 2 |\alpha| \bra \cos(\phi-\Phi)
\ket_{P(\phi)}$.

\begin{figure}
\center{\epsfig{figure=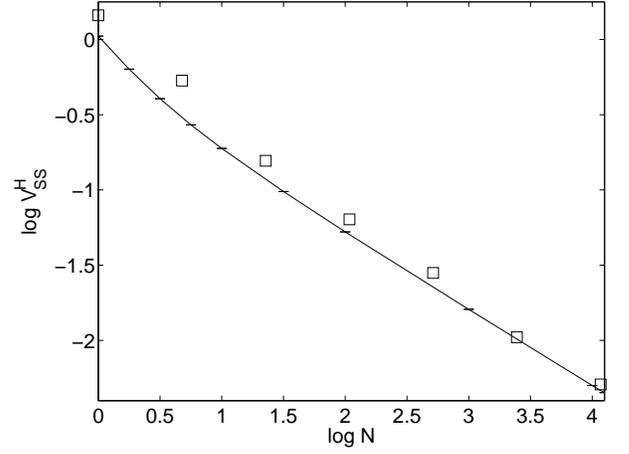,width=80mm}} \caption{Log-log
plots of the logarithm of steady-state Holevo variance
$V^{H}_{SS}$ versus the photon flux $N$ for the BW adaptive
(squares) and the semi-optimal adaptive (solid line)
phase-estimation schemes. Both $V^{H}_{SS}$ and $N$ are
dimensionless.} \label{graph_two}
\end{figure}

To obtain $V^{H}_{SS}$ from Eq.~(\ref{kse_semi}) we applied the
same method used for the optimal heterodyne-based scheme centred
around decomposing $P(\phi)$ via the Fourier decomposition in
Eq.~(\ref{fourier_decomp}). The results obtained are plotted in
Fig.~\ref{graph_two}. In addition, for small and large $N$ the
following analytical results were found:
\begin{equation}
\;\;\;\;\;\;V^{H}_{SS} \simeq 1/ (2\sqrt{N}) \;\;({\rm large}\;N)
\end{equation}
\begin{equation} \label{small_N_adaptive}
V^{H}_{SS} \simeq 1/N \;\;({\rm small}\;N).
\end{equation}
These results were obtaining via calculations very similar to
those used in Sect.~\ref{opt_het_subsub} to obtain the
corresponding estimates for optimal heterodyne detection.

\begin{figure}
\center{\epsfig{figure=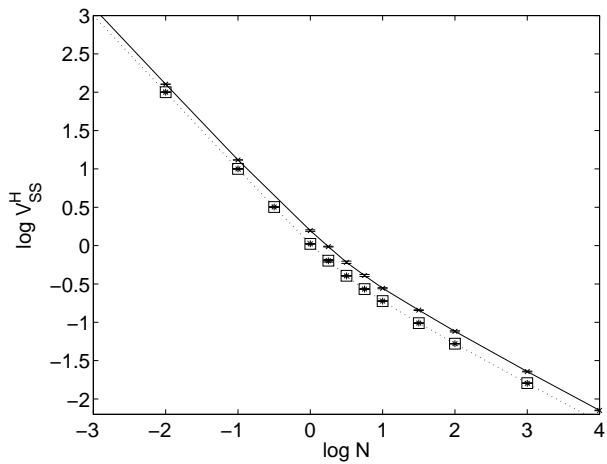,width=80mm}} \caption{Log-log
plots of the logarithm of steady-state Holevo variance
$V^{H}_{SS}$ versus the photon flux $N$ for the optimal
heterodyne-based (solid line), semi-optimal adaptive (squares) and
simple adaptive phase-estimation schemes (dotted line with *'s).
The large $N$ and small $N$ results lie upon the asymptotes
derived for these regions. Both $V^{H}_{SS}$ and $N$ are
dimensionless.} \label{graph_three}
\end{figure}

\subsubsection{Comparison}
Figs~\ref{graph_two} and \ref{graph_three} display a number of
interesting features which we now highlight. First,
Fig.~\ref{graph_three} shows that the semi-optimal adaptive and
simple adaptive schemes are equally accurate, as evidenced by the
fact that they have identical $V_{SS}^{H}$-versus-$N$ plots.
Second, Fig.~\ref{graph_two} illustrates that the semi-optimal
adaptive scheme (and hence also the simple adaptive scheme) is
more accurate than BW adaptive one for all $N$ values except when
$N \gtrsim 10^{3.5}$. Third, Fig.~\ref{graph_three} demonstrates
that the semi-optimal adaptive scheme is significantly more
accurate than the optimal heterodyne-based or canonical schemes.
Fourth, Fig.~\ref{graph_three} also shows that adaptive phase
estimation is more accurate than {\em any} non-adaptive
phase-estimation scheme in which the field is measured in real
time. The reason for this is the following: Assume that we measure
the field non-adaptively in real time. By this we mean that we
measure it via a continuous sequence of identical
infinitesimal-time measurements and thus measure each spatial
`segment' of the signal beam as it is incident on the detector. In
this scenario, the best measurement we can make is a canonical
phase measurement (as we must decide what to measure whilst
knowing nothing about the phase). However, from
Subsection~\ref{theory_canonical}, we know that estimating $\phi$
non-adaptively via such a measurement leads to an estimate only as
accurate as that of the optimal heterodyne-based scheme. We also
know that adaptive phase-estimation is more accurate this latter
non-adaptive scheme in the cw scenario and hence it is also more
accurate than the canonical non-adaptive scheme.

%It thus follows that in this scenario adaptive phase estimation is
%more accurate {\em any} non-adaptive phase estimation scheme in
%which the field is measured in real time. This is one of our two
%main results. It shows that we can estimate $\phi$ more accurately
%in the scenario considered using adaptive phase estimation than
%via a wide class of non-adaptive estimation schemes including the
%conventional one, namely one based on real-time heterodyne
%detection.
%This is because, given that the
%scheme is non-adaptive, we
%Canonical is, most
%probably, the best you could do non-adaptively assuming that we
%restrict ourselves to "Markovian" non-adaptive schemes. That is,
%we measure every bit of light as soon as it enters the detector
%region. Thus, adaptive is, most probably, better than any
%non-adaptive scheme (assuming no non-Markovian non-adaptive
%schemes could do better). We can't be 100 \%, absolutely sure that
%adaptive is better than the absolute very best non-adaptive,
%heterodyne-based or otherwise.

%({\em Show graphs comparing the estimates. 1. BW het, BW adapt 2.
%semi-optimal adapt/ simple adapt, opt. het./canonical}
%{\em -put all asymptotics on the graphs, small and large.})

\section{Discussion} \label{discussion}

%
%why isn't canonical the best?
%

The results of Section~\ref{results_sec} display a number of
interesting features which we now discuss. First, it might seem
puzzling that the canonical phase-estimation scheme is only as
accurate as the optimal heterodyne-based scheme and is not,
instead, the most accurate scheme. Given that a canonical phase
measurement is generally thought to be the best measurement of
phase we can make, why isn't the canonical scheme the most
accurate? The answer to this lies in the details of the scenario
we consider. In the standard scenario in which we wish to estimate
phase, we make a single phase measurement on a system for which we
have no prior information about the phase. In this scenario, a
canonical measurement is optimal. However, in the scenario we
consider prior to making a measurement on the field at time $t
\neq 0$, we already know something about $\phi$, as evidenced by
the fact we possess a non-trivial probability distribution
$P(\phi)$. This prior information can be exploited by measurements
other than a canonical one to yield more information about phase
than would a canonical measurement.

To understand the preceeding point it may be helpful to consider
the following example: Say we wish to determine as accurately as
possible the phase of a system in a weak coherent state which we
know to be either one of the two states $| \psi_{\pm} \ket= |0\ket
+  \gamma e^{\pm i \phi} |1 \ket$, where $\gamma \in \mathbb{R}
\ll 1$, with equal probability. In this instance, because we
already know something about $\phi$, we can tailor the measurement
in accordance with this prior knowledge and measure the
$\Phi=\pi/2$ or $Y$ quadrature to obtain slightly more information
about $\phi$ than would a canonical measurement. Specifically,
measuring the $Y$ quadrature, we estimate $\phi$ correctly with
probability $1/2 + 0.799 \gamma \sin \phi$, while for a canonical
measurement this probability is only $1/2+ 0.638 \gamma \sin \phi
$.

%
%we are uncertain if adaptive is better than any non-adaptive
%

Another interesting feature related to Section~\ref{results_sec}'s
results concerns the main conclusion we drew from them, which was
that adaptive phase estimation in the cw scenario is more accurate
than any non-adaptive scheme in which the field is measured in
real time. Though we were able to arrive at this result, we are
uncertain if adaptive phase estimation is better than any
non-adaptive scheme at all. This is because it is conceivable that
there exists a non-adaptive scheme in which, instead of measuring
the field in real time, we store up a portion of it over a period
of time and then measure the accumulated field as a whole that is
more accurate than adaptive phase estimation.

%
%why does simple adaptive do so well?
%

The results of Section~\ref{results_sec} also show that simple
adaptive does as well as semi-optimal adaptive. Why does this
relatively uncomplicated scheme do so well? One possibility is
that the state of the beam we consider, being based on coherent
states, is somewhat 'simple'. Perhaps, it does not allow us to
fully utilise the power of the more complicated semi-optimal
adaptive scheme.

%
%experiments
%

One final interesting feature of Section~\ref{results_sec}'s
results concerns the variation with $N$ of the relative
superiority of adaptive phase estimation over real-time
nonadaptive phase estimation. This is measured by the ratio of the
steady-states Holevo variances for the optimal heterodyne-based
and the semi-optimal adaptive schemes. For $N \ll 1$, this ratio
is given by Eqs~(\ref{small_N_het}) and (\ref{small_N_adaptive})
and is $4/\pi \simeq 1.27$ while for $N \gg 1$ it is $\sqrt{2}
\simeq 1.41$. For intermediate $N$ values, it lies in between
these two extremes. Of particular importance is the fact that the
gap is present for $N \simeq 1$. This is because this regime is
the most fertile for experimental implementation as within it the
errors we wish to see are not swamped by technical noise. It is
also noteworthy that the small-$N$ ratio of $4/\pi$ is
significantly greater than the analogous ratio in Ref.
\cite{berry02}, which was approximately 1.1, between the adaptive
and non-adaptive estimates in this other paper.

%
%is semi-optimal adaptive the best adaptive scheme?
%

%The semi-optimal adaptive scheme is quite accurate, but is it the
%most accurate adaptive scheme? We are unsure of this and it is
%possible that one may more accurately estimate phase by
%occasionally measuring the amplitude quadrature so as to try to
%check that our estimate is not out by $\pi$.

%
%wider implications of the results
%

%This paper's main result fits in well with a number of others
%showing the usefulness of feedback at the quantum level (which are
%too numerous to cite). This type of feedback is known as quantum
%feedback or (more accurately) quantum-limited feedback. The
%results all concern how accurately we can measure quantum systems
%in the face of intrinsic quantum noise. Furthermore, they often do
%this with a view to controlling these systems.

%
%relation to deficiency in squeezed-beam calculation
%

Having discussed the results in Section~\ref{results_sec}, we now
turn to two theoretical issues arising from our work. First, in
this paper we have considered estimating the phase of an EM beam
in a coherent state. However, other beams could be investigated as
was done in Ref.~\cite{berry02} which looked at a so-called
squeezed EM beam with a randomly fluctuating phase. That paper
found that, for such a beam, adaptive phase estimation was more
accurate than heterodyne-based non-adaptive phase-estimation not
just by a constant factor (as this paper has), but by a factor
scaling with $N$. In particular, it found that for such a beam the
steady-state Holevo variance of the error scaled as $N^{-2/3}$ in
adaptive phase-estimation but only as $N^{-1/2}$ in
heterodyne-based non-adaptive phase estimation.

While this result for squeezed beams is interesting, the
calculations behind it contained a number of deficiencies. First,
Ref.~\cite{berry02} considered a beam with broadband squeezing,
i.e. one that was squeezed at all frequencies, and thus the noise
present in the beam had infinite energy. The parameter
$N=|\alpha|^{2}/\kappa$ was finite, however, as it relates only to
the energy carried by the mean field. Such a beam is unphysical
and, furthermore, constitutes an inappropriate theoretical model
for the problem considered, as we shall soon see. The second
deficiency in the calculation was that it involved estimating
$\phi(t)$ using only information about the beam's signal. This
meant that information in the beam's noise was ignored. If such
information had been used then, as the noise had infinite energy,
we could have instantly determined $\phi$ by determining the
relative sizes of the noise in different quadratures. Thus, the
calculation in Ref.~\cite{berry02} ignored obtaining phase
information from a potential source (the noise) and revolved
around a model such that if we do consider this potential source,
we find that we can instantly determine $\phi(t)$ with perfect
accuracy, which is unrealistic.
%They estimated $\phi$ from just the signal and ignored all the
%information in the noise which would have allowed them to
%instantaneously determine $\phi$.
Because of these deficiencies, we feel that it is desirable to do
additional calculations on squeezed beams. We anticipate that our
`optimal' approach to obtaining phase estimates based on quantum
parameter estimation may be useful in such calculations.

%
%even if we don't know |\alpha| precisely, we can do OK
%

A second theoretical issue arising from our work is the following:
Throughout the paper, it was assumed that $|\alpha|$ was known
precisely. However, even if we only know that $|\alpha| \geq a$,
where $a \in \mathbb{R}$ we can still do at least as well as when
we know that it equals $a$. This follows on from work by Stockton
{\em et al.} \cite{doherty03} (Sect. V). Knowing $|\alpha|$
precisely, we have, for the simple adaptive (and semi-optimal
adaptive) schemes,
\begin{eqnarray}
\chi_{\rm opt} = 2\sqrt{\kappa} |\alpha|.
\end{eqnarray}
If we only know that $|\alpha| \geq a$ we can set $\chi$ equal to
\begin{eqnarray}
\chi = 2\sqrt{\kappa} a.
\end{eqnarray}
For $N \gg 1$, this leads to \cite{berry02}
\begin{eqnarray}
V^{H}_{SS} & = & 2 \sqrt{\kappa} a/(8\alpha^{2}) + {\sqrt
\kappa}/(2a) \\
& \simeq & {\sqrt \kappa}/(2a).
\end{eqnarray}
That is, we can estimate $\phi$ at least as well as we can
assuming we know that $|\alpha|$ is exactly the minimum known
value.

%
%squeezed beam work of Berry and Wiseman doesn't work.
%

%Though we considered a coherent beam, previous work has looked
%squeezed beams \cite{berry02} in the cw scenario. However, this
%work contained some deficiencies. First, it assumed that the
%squeezing was broadband with a flat spectrum. Because of this the
%beam's photon flux was infinite and thus unphysical. Furthermore,
%the phase estimate found in \cite{berry02} was calculated only by
%using information in the signal component of the measured current.
%As a consequence, it ignored information about $\phi$ contained in
%the noise. For a squeezed beam the relative sizes of the noises of
%different quadratures yields information about $\phi$.
% which
%can tell one info. about $\phi$ for a squeezed beam. Taking this
%into account, we would determine $\phi$ instantly.
%Given these things, a calculation avoiding these deficiencies
%could be carried out. We expect that the theoretical technique for
%estimating $\phi$ based on quantum parameter estimation introduced
%in Subsection~\ref{parameter_estimation_theory} will be useful in
%such a calculation.

\section{Conclusion}
Quantum phase estimation and, in particular, Bayes' rule were used
to find optimally accurate phase estimates and to show that, for a
continuous EM beam with a randomly fluctuating phase, adaptive
phase-estimation is more accurate than any non-adaptive
phase-estimation scheme in which the field is measured in real
time. Though it is more accurate for all photon fluxes it is, in
particular, more accurate for such beams possessing small to
moderate photon fluxes. This is important as this is the regime in
which in experiments would have the greatest chance of confirming
any theoretical difference between the two types of
phase-estimation schemes.

\section{Appendix A}
This section details the known correspondence between a Zakai
equation of the form
\begin{equation} \label{general_zakai}
d\tilde{P} =  (X I + {\rm c.c.}) \tilde{P} dt
\end{equation}
and the KS equation
\begin{equation} \label{general_kse}
dP = \left[(X-\bra X \ket_{P}) (I -\bra I \ket_{P}) + {\rm c.c.}
\right] P dt.
\end{equation}
To obtain Eq.~(\ref{general_kse}) from Eq.~(\ref{general_zakai}),
we begin with the identity
\begin{eqnarray}
P(\phi)+ dP(\phi) = \frac{\tilde{P} + d\tilde{P}}{\int_{\phi}
d\phi \tilde{P} + d\tilde{P}}.
\end{eqnarray}
Taking out a factor of $\int_{\phi}d\phi {\tilde P}(\phi)$ in the
denominator leads to
\begin{equation}
P(\phi)+ dP(\phi) = \frac{\tilde{P} + d\tilde{P}} {\int_{\phi}
d\phi \tilde{P} \left[ 1+ \frac{1}{\int_{\phi} d\phi \tilde{P}}
\int_{\phi} d\phi  d\tilde{P} \right]}
\end{equation}
Expanding the expression in the denominator within the square
pararentheses as a power series using the binomial theorem
($(1+x)^{n}=1+nx + n(n-1) x^{2}/2 + \ldots$), yields
\begin{equation}
P(\phi)+ dP(\phi) \simeq \frac{\tilde{P} + d\tilde{P}}
{\int_{\phi} d\phi \tilde{P}} \left[ 1- \frac{\int_{\phi} d\phi
d\tilde{P}}{\int_{\phi} d\phi \tilde{P}} + \frac{ \left(
\int_{\phi} d\phi d\tilde{P} \right)^{2}}{\left( \int_{\phi} d\phi
\tilde{P} \right)^{2} } \right].
\end{equation}
Normalising the distribution $\tilde{P}$ using the factors of
$\int_{\phi} d\phi P(\phi)$ in the denominator and also
substituting in the expression for $dP$ in
Eq.~(\ref{general_zakai}), we obtain
\begin{eqnarray} \nonumber
P + dP & = & \left( P + (X I + {\rm c.c.}) P dt \right)
\\ \nonumber & &  \times \left( 1 - \int_{\phi} d\phi (X I + {\rm c.c.}
)P(\phi) dt \right.
\\ & & \left. + \left( \int_{\phi} d\phi (X I + {\rm c.c.} )P(\phi) dt
\right)^{2} \right) \\ \nonumber & = & \left( P + (X I + {\rm
c.c.}) P dt \right) \left( 1 \right.
\\ \nonumber
& &  \left. -( \bra X \ket_{P} I + {\rm c.c.}) dt + \left( \bra X
\ket_{P} I + {\rm c.c.} ) dt
\right)^{2} \right). \\
\end{eqnarray}
Expanding this expression and keeping only terms of order $dt$ or
less, we arrive at Eq.~(\ref{general_kse}).
%\begin{eqnarray} \nonumber
%dP & = & \left[(X-\bra X \ket) (I -\bra I \ket) + {\rm c.c.} \ket
%\right] P dt
%%\left( P + (X I + {\rm c.c.}) P dt \right)
%%\\ \nonumber
%& &  \times \left( 1 - ( \bra X \ket I + {\rm c.c.}) dt + \left(
%\bra X \ket I + {\rm c.c.} ) dt
%\right)^{2} \right). \\
%\end{eqnarray}
%{\em Howard --- I can't quite get
%this calculation to work. I'm having some trouble with $dt^{2}$
%terms that get converted to $dt$ terms using the identity
%$dWdW^{*} = dt$.}
\section{Appendix B}
In this appendix we demonstrate that, for the schemes based on
quantum parameter estimation (the optimal heterodyne-based, the
canonical and semi-optimal adaptive schemes),
\begin{equation} \label{identity_one}
\bra e^{i(\phi -{\hat \phi})} \ket_{\xi,I} = \bra |\bra e^{i \phi}
\ket_{P(\phi)} | \ket_{I}.
\end{equation}
By definition
\begin{equation} \label{identity_two}
\bra e^{i(\phi -{\hat \phi})} \ket_{\xi,I} = \int_{\xi} \int_{I}
d\xi dI P(\xi, I) e^{i(\phi(\xi) -{\hat \phi}(I))}.
\end{equation}
Expressing $e^{i(\phi(\xi) -{\hat \phi}(I))}$ as an integral over
the dummy phase variable $\varphi$, we obtain
\begin{eqnarray} \label{identity_three}
e^{i(\phi(\xi) -{\hat \phi}(I))} &= &\int_{\varphi} d\varphi
\delta(\phi(\xi) - \varphi) e^{i[\varphi - {\hat \phi}(I)]}.
\end{eqnarray}
Substituting the right-hand side of Eq.~(\ref{identity_three})
into the right-hand side of Eq.~(\ref{identity_two}) yields
\begin{eqnarray} \label{identity_three_a} \nonumber
\bra e^{i(\phi -{\hat \phi})} \ket_{\xi,I}= \;\;\;\;\;\;\;\;\;
\;\;\;\;\;\;\;\;\;\;\;\;\;\;\;\;\;\;\;\;\;\;\;\;\;\;\;\;\;\;\;\;\;\;\;\;\;\;\;\;\;\;
& & \\ \nonumber \int_{\xi} \int_{I} \int_{\varphi} d\xi dI
d\varphi P(\xi) P(I | \xi) \delta(\phi(\xi)
- \varphi) e^{i[\varphi - {\hat \phi}(I)]}. & & \\
\end{eqnarray}

Assuming we know the so-called process noise $\xi$, then we know
the phase $\phi$ exactly and thus our probability density function
for $\phi$ is a Dirac delta. From this it follows that
\begin{equation}
P(I | \xi) \delta(\phi(\xi) - \varphi)d\varphi =P(\varphi,I | \xi)
d\varphi.
\end{equation}
Substituting this result into Eq.~(\ref{identity_three_a}) and
integrating over $\xi$ yields
\begin{eqnarray} \label{identity_three_b} \nonumber
\bra e^{i(\phi -{\hat \phi})} \ket_{\xi,I} & = & \int_{I}
\int_{\varphi} dI d\varphi
 P(\varphi,I) e^{i[\varphi - {\hat
\phi}(I)]}. \\
\end{eqnarray}
Using elementary probability theory, we obtain
\begin{eqnarray} \label{identity_three_b_two} \nonumber
\bra e^{i(\phi -{\hat \phi})} \ket_{\xi,I} & = &
 \int_{I} dI P(I) \int_{\varphi} d\varphi P(\varphi|I) e^{i[\varphi - {\hat
\phi}(I)]}. \\
\end{eqnarray}
Given that
\begin{equation}
{\hat \phi}(I) = {\rm arg} \left( \int d\varphi' P(\varphi' | I)
e^{i \varphi'} \right),
\end{equation}
where $\varphi'$ is a second dummy phase variable,
Eq.~(\ref{identity_three_b_two}) leads to
\begin{eqnarray} \label{identity_three_c} \nonumber
\bra e^{i(\phi -{\hat \phi})} \ket_{\xi,I} & = & \int_{I} dI P(I)
\left| \int_{\varphi} d\varphi P(\varphi|I) e^{i \varphi} \right| \\
& = & \bra | \bra  e^{i \varphi}  \ket_{P(\varphi)} | \ket_{I}.
\end{eqnarray}
Upon replacing $\varphi$ by $\phi$ in the final expression, where
$\phi$ now acts as a dummy phase variable,
Eq.~(\ref{identity_one}) is obtained.

\section{Acknowledgements}
This work was supported by the Australian Research Council. DTP
would like to thank Drs Dominic Berry and Kurt Jacobs, and Mr Neil
Oxtoby, for their assistance.

\vspace*{20cm}

\begin{table}[h]
\hspace*{3cm} \centerline{
\begin{tabular}{|c|c|c|c|} \hline
Name of Measurement Scheme & ${\hat \phi}$ & $d\Phi/dt$ &  Type of Detection \\
\hline \hline
Canonical & ${\rm arg} \left( \bra
\exp(i\phi) \ket_{P(\phi)} \right)$ & N/A &  canonical \\
\hline Optimal heterodyne-based & ${\rm arg} \left( \bra
\exp(i\phi) \ket_{P(\phi)} \right)$
& $\Delta$ & heterodyne \\
\hline BW heterodyne-based & ${\rm arg} \left(A_{t} \right)$ & $\Delta$ & heterodyne \\
\hline  \hline
BW adaptive & ${\rm arg} \left(A_{t} + \chi B_{t}A_{t}^{*} \right)$ & $\sqrt{\kappa} I_{\rm r}$ &  homodyne \\
\hline semi-optimal adaptive & ${\rm arg} \left( \bra
\exp(i\phi) \ket_{P(\phi)} \right)$ & $d{\hat \phi}/dt$ & homodyne \\
\hline simple adaptive & ${\rm arg} \left( A_{t} \right)$ &
$\sqrt{\kappa} I_{\rm r}$ & homodyne \\ \hline
\end{tabular}}
\caption{Summary of phase estimates} \label{table}
\end{table}


\begin{thebibliography}{99}
\bibitem{hall91} M. J. Hall and I. G. Fuss, Quantum Opt. {\bf 3},
147 (1991).
\bibitem{caves94} C. M. Caves and P. D. Drummond, Rev. Mod. Phys.
{\bf 66}, 481 (1994).
\bibitem{mabuchi04} H. Mabuchi, private communication.
\bibitem{pegg88} D.T. Pegg and S.M. Barnett, Europhys. Lett. {\bf 6}, 483
(1988).
\bibitem{verstraete01} S. L. Braunstein, A. S. Lane, and C. M.
Caves, Phys. Rev. Lett. {\bf 69}, 213 (1992); B. C. Sanders and G.
J. Milburn, Phys. Rev. Lett. {\bf 75}, 2944 (1995); H. Mabuchi,
Quant. Semiclass. Opt. {\bf 8}, 1103 (1996); F. Verstraete, A. C.
Doherty, and H. Mabuchi, Phys. Rev. A {\bf 64}, 032111 (2001); P.
Warszawski, J. Gambetta, and H. M. Wiseman, Phys. Rev. A {\bf 69},
042104 (2004).
\bibitem{wiseman95} H. W. Wiseman, Phys. Rev. Lett. {\bf 75}, 4587 (1995).
\bibitem{wiseman97} H. M. Wiseman and R. B. Killip, Phys. Rev. A {\bf 56}, 944
(1997).
\bibitem{wiseman98} H. M. Wiseman and R. B. Killip, {\bf 57}, 2169 (1998).
\bibitem{armen02} M. A. Armen {\em et al.}, Phys. Rev. Lett. {\bf 89}, 133602
(2002).
\bibitem{berry02} D. Berry and H. M. Wiseman, Phys. Rev. A {\bf 65}, 043803 (2002).
\bibitem{gardiner83} C. W. Gardiner {\em Handbook of stochastic methods for physics, chemistry, and the natural
sciences} (New York, N.Y., Springer-Verlag, 1983), Chapter 4.
\bibitem{yuen78} H. P. Yuen and J. H.
Shapiro, IEEE Trans. Inf. Theory {\bf IT-24}, 675 (1978).
\bibitem{yuen80} H. P. Yuen and J. H.
Shapiro, IEEE Trans. Inf. Theory {\bf IT-26}, 78 (1978).
\bibitem{shapiro79} J. H. Shapiro, H. P. Yuen, and J. A. Machado
Mata, IEEE Trans. Inf. Theory {\bf IT-25}, 179 (1979).
\bibitem{shapiro84} J. H.
Shapiro and S. S. Wagner, IEEE Quantum Electron. {\bf QE-20}, 803
(1984).
\bibitem{shapiro85} J. H. Shapiro, IEEE Quantum Electron. {\bf QE-21}, 237 (1985).
\bibitem{mandel} L. Mandel and E. Wolf, {\em Optical Coherence and Quantum
Optics} (Cambridge University Press, Cambridge, 1995).
\bibitem{barnett93}S.M. Barnett and B. J. Dalton, Physica Scripta {\bf T48}, 13
(1993).
\bibitem{yariv89} A. Yariv, {\em Quantum Electronics}, (John Wiley \& Sons, New York, 1989).
\bibitem{holevo84} A. S. Holevo, Lect. Notes Math. {\bf 1055}, 153 (1994).
\bibitem{benssousan92} {\em Stochastic Control of Partially
Observable Systems} (Cambridge University Press, Cambridge, 1992).
\bibitem{mcgarty74} T. P. McGarty, {\em Stochastic Systems and State Estimation}
(Addison-Wesley, Sydney, 1974).
\bibitem{doherty00} A. Doherty {\em et al.}, Phys Rev. A {\bf 62},
012015 (2000).
\bibitem{wiseman93} H. M. Wiseman, Phys. Rev. A {\bf 47}, 5180
(1993).
\bibitem{leonhardt95} U. Leonhardt, J. A. Vaccaro, B. B\"{o}hmer, and
H. Paul Phys. Rev. A {\bf 51}, 84 (1995).
\bibitem{doherty03} J. K. Stockton, JM Gerimia, A. C. Doherty, and H. Mabuchi,
Phys. Rev. A {\bf 69}, 032109 (2004).
\end{thebibliography}
\end{document}